\documentclass[useAMS,usenatbib]{mn2e}
\usepackage[utf8]{inputenc}
\usepackage{epsfig,graphicx,latexsym,amsmath,amssymb}
\usepackage{natbib}
\usepackage[draft]{hyperref}
\usepackage{url}
\usepackage{mathrsfs}
\usepackage{lastpage}
\usepackage{color}
\usepackage{colortbl}
\usepackage{xcolor}
\usepackage{listings}
\usepackage{bm}
\usepackage{float}
\usepackage{multicol}

\usepackage{tabularx}
\definecolor{gray97}{gray}{.97}
\definecolor{gray75}{gray}{.75}
\definecolor{gray45}{gray}{.45}

\newcommand{\nbody}{$N-$body}
\newcommand{\GR}{{\sc GraviDy}}
\newcommand{\nbg}{{NBODY6GPU}}
\newcommand{\bs}[1]{\boldsymbol{\mathbf{#1}}}

\lstnewenvironment{listing}[1][]{\lstset{#1}\pagebreak[0]}{\pagebreak[0]}
\lstdefinestyle{consola}{basicstyle=\scriptsize\bf\ttfamily,
                         backgroundcolor=\color{gray75},
                        }

\bibpunct[,]{(}{)}{;}{a}{}{,}

\title[\GR: Dynamics with softening]{GraviDy, a GPU modular, parallel
    direct-summation $N-$body integrator: Dynamics with softening}

\author[C. Maureira-Fredes \& P. Amaro-Seoane]
{Cristián Maureira-Fredes$^{1,\,2}$
                        \thanks{E-mail: Cristian.Maureira.Fredes@aei.mpg.de (CM)} \&
Pau Amaro-Seoane$^{3,\,4\,5}$\\
$^{1}$
Max Planck Institute for Gravitational Physics
(Albert-Einstein-Institut), D-14476 Potsdam, Germany.
\\
$^{2}$
Universidad Técnica Federico Santa María,
Avenida España 1680, Valparaíso, Chile.
\\
$^{3}$
Institut de Ci{\`e}ncies de l'Espai (CSIC-IEEC) at Campus UAB, Carrer de Can Magrans s/n 08193 Barcelona, Spain.\\
$^{4}$
Institute of Applied Mathematics, Academy of Mathematics and Systems Science, CAS, Beijing 100190, China.\\
$^{5}$
Kavli Institute for Astronomy and Astrophysics, Beijing 100871, China.
}

\begin{document}

\date{draft \today}
\maketitle
\label{firstpage}

\begin{abstract}
A wide variety of outstanding problems in astrophysics involve the motion of a large
number of particles under the force of gravity.
These include the global evolution of globular clusters, tidal disruptions of stars
by a massive black hole, the formation of protoplanets and sources of gravitational
radiation.
The direct-summation of $N$ gravitational forces is a complex problem with no
analytical solution and can only be tackled with approximations and numerical
methods.
To this end, the Hermite scheme is a widely used integration method.
With different numerical techniques and special-purpose hardware, it can be used to
speed up the calculations.
But these methods tend to be computationally slow and cumbersome to work with.
We present a new GPU, direct-summation $N-$body integrator written from scratch and
based on this scheme, which includes relativistic corrections for sources of
gravitational radiation.
GraviDy has high modularity, allowing users to readily introduce new physics, it
exploits available computational resources and will be maintained by regular updates.
GraviDy can be used in parallel on multiple CPUs and GPUs, with a considerable
speed-up benefit.
The single GPU version is between one and two orders of magnitude faster than the
single CPU version.
A test run using 4 GPUs in parallel shows a speed up factor of about 3 as compared
to the single GPU version.
The conception and design of this first release is aimed at users with access to
traditional parallel CPU clusters or computational nodes with one or a few GPU cards.
\end{abstract}

\begin{keywords}
Methods: numerical -- Stars: Kinematics and dynamics -- Celestial Mechanics

\end{keywords}

\section{Motivation}
\label{sec.motivation}

The dynamical evolution of a dense stellar system such as e.g. a globular
cluster or a galactic nucleus has been addressed extensively by a number of
authors. For Newtonian systems consisting of more than two stars we must rely
on numerical approaches which provide us with solutions that are more or less
accurate. In this sense, one could make the following coarse categorisation of
integration schemes for pure stellar dynamics: those which are particle-based
and those which are not. In the latter, the system is treated as a continuum,
so that while we know the general properties of the stellar system such as the
mean stellar density, of the average velocity dispersion, we do not have
information about specific orbits of stars. To this group, belongs direct
integration of the Fokker-Planck equation \citep{IW84,KLG98} or moments of it
\citep{ASEtAl04,SchneiderEtAl11}, including Monte Carlo approaches to the
numerical integration of this equation \citep{SH71b}. A particle-based
algorithm, however, assumes that a particle is tracing a star, or a group of
them. In this group, the techniques go back to the early 40's and involved
light bulbs \citep{Holmberg1941}.  The first computer simulations were
performed at the Astronomisches Rechen Institut, in Heidelberg, Germany, by
\citep{vonHoerner1960,vonHoerner1963}, using 16 and 25 particles. These first
steps led to the modern $N-$body algorithms.

We can distinguish two types of {\nbody} algorithms: the so-called
collision-less, where a star just sees the background potential of the rest of
the stellar system \citep[e.g. the Barnes-Hut treecode or the fast multipole
method][ which scale as $O(N\log N)$ and $O(N)$, with $N$ the particle number,
respectively]{BarnesHut86,GreendardThesis}, and the more expensive collisional
one, or ``direct-summation'', in which one integrates all gravitational forces
for all stars to take into account the graininess of the potential and
individual time steps, to avoid large numerical errors.  This is important in
situations in which close encounters between stars play a crucial role, such as
in galactic nuclei and globular clusters, because of the exchange of energy and
angular momentum. The price to pay however is that they typically scale as
$O(N^{2})$.

A very well known example is the family of direct-summation {\sc Nbody}
integrators of Aarseth \citep[see
e.g.][]{Aarseth99,Spurzem1999,Aarseth03}\footnote{All versions of the code are
publicly available at the URL\\
\url{http://www.ast.cam.ac.uk/~sverre/web/pages/nbody.htm}} or also {\sc kira}
\citep[see][]{PortegiesZwartEtAl01}\footnote{\url{http://www.sns.ias.edu/~starlab/}}.
The progress in both software and hardware has reach a position in which
we start to get closer and closer to simulate realistic systems.

However, the scaling $O(N^{2})$ requires supercomputers, such as traditional
Beowulf clusters, which requires a parallelisation of the code, such as the
version of {\sc Nbody6} developed by Spurzem and collaborators, {\sc
Nbody6++}\footnote{Available at this URL
\url{http://silkroad.bao.ac.cn/nb6mpi}} \citep{Spurzem1999}, or special-purpose
hardware, like the GRAPE (short for GRAvity
PipE\footnote{\url{http://grape.c.u-tokyo.ac.jp/grape}}) system. The principle
behind GRAPE systems is to run on a special-purpose chip the most time
consuming part of an $N-$body simulation: the calculation of the accelerations
between the particles. The remainder is calculated on a normal computer which
serves as host to the accelerator board(s) containing the special purpose
chips. Such a system achieves similar or even higher speeds than
implementations of the $N-$body problem on supercomputers \citep[see
e.g.][]{TMFES96,MT98,Makino98,GRAPE6A}.

On the other hand, modern graphics processing units (GPUs) offer a very
interesting alternative. They have been mostly used in game consoles, embedded
systems and mobile phones. They were originally used to perform calculations
related to 3D computer graphics.  Nevertheless, due to their highly parallel
structure and computational speed, they can very efficiently be used for
complex algorithms.  This involves dealing with the parallel computing
architecture developed by NVIDIA\footnote{\url{http://www.nvidia.com}}, the
Compute Unified Device Architecture (CUDA).This is the main engine in NVIDIA
GPUs, and it has been made accessible to developers via standard programming
languages, such as C with NVIDIA extensions compiled thanks to a PathScale
Open64 C compiler.
This is what allows us to create binary modules to be run on the
GPUs.
Another option is Open Computing Language
(OpenCL)\footnote{\url{https://www.khronos.org/opencl/}},
which offers a framework to write parallel programmes for heterogeneous systems,
including also computational nodes with field-programmable gate arrays (FPGAs),
digital signal processors (DSPs), among others.
CUDA, and also OpenCL are  ``the doors'' to the native instruction set and memory
of the parallel elements in the GPUs.
This means that these can be handled as open
architectures like CPUs with the enormous advantage of having a parallel-cores
configuration. More remarkably, each core can run {\em thousands} of processes
at the same time.
We selected CUDA over OpenCL, because our systems are equipped with NVIDIA GPUs,
even though we note that
OpenCL has shown similar performance to CUDA in $N-$body
simulations~\citep{CapuzzoDolcettaSpera2013}.

There has been recently an effort at porting existing codes to this
architecture, like e.g. the work of
\cite{Portegies2007a,Hamada2007,Belleman2008} on single nodes or using large
GPU clusters
\citep{berczik2011high,NitadoriAarseth2012,Capuzzo-DolcettaEtAl2013} and
recently, the work by \cite{bercziketal2013} using up to 700 thousand GPU
cores for a few million bodies simulation with the
$\phi-$GPU~\footnote{\url{ftp://ftp.mao.kiev.ua/pub/berczik/phi-GPU/}} code,
which reached in their work about the half of the peak of the new Nvidia Kepler
K20 cards.

Large-scale (meaning number of particles) simulations have recently seen an
important improvement with the work of \cite{WangEtAl2015,WangEtAl2016}. In his
more recent work of 2016, Wang and collaborators integrated systems of one
million bodies in a globular cluster simulation, using from ~2,000 to ~8,600
hours of computing time.\footnote{This
impressive achievement was rewarded with a bottle of Scotch whisky (not
whiskey), kindly and generously offered to him by Douglas Heggie during the
excellent MODEST 15-S in Kobe.}

In this paper we present the initial version of {\GR} ({\sc
Gr}a{\sc{vi}}tational {\sc dy}namics), a highly-modular, direct-summation
{\nbody} code written from scratch using GPU technology ready to integrate a
pure dynamical gravitational system.  In section~\ref{sec.algorithm} we present
in detail the structure of the code, the most relevant and innovative parts of
the algorithm, and their implementation of the scheme in the idiom of GPU
computing. In section~\ref{sec.experiments} we check our code with a series of
well-known tests of stellar dynamics for a dense stellar system and evaluate
global dynamical quantities and we also evaluate the performance of the GPU
version against the CPU one. In section~\ref{sec.PN} we present the implementation
of the relativistic corrections, and a set of tests.
In section~\ref{sec.conclusions} we summarise our work and give a short description
of the immediate goals that will be described
in upcoming publications.

\textit{We have decided to focus on single-node clusters (meaning one or more
GPU cards embedded in a host PC) and traditional multi-CPU clusters (e.g.
Beowulf clusters), since this setup is more common to most users who aim to
run middle-scale simulations. In the appendices we give a succinct description
on how to download the code, how to compile it, and the structure of the data.
We also include a set of python tools to analyse the
results. Moreover, we also introduce a simple visualisation tool based on {\tt
OpenGL}, which can provide us with information sometimes difficult to obtain
with two-dimensional graphics.  In particular, we have made a significant
effort in documentation and modularity, since it is our wish that the code is
used, shaped and modified at will.}


\section{The current algorithm}
\label{sec.algorithm}

\subsection{The integration scheme}
\label{sec:hermite}

In this section we give a very brief introduction to the numerical {\nbody}
problem.  We refer the reader to e.g. \cite{Aarseth03,HeggieHut03} or the
excellent on-line course ``The art of computational
science''\footnote{\url{http://www.artcompsci.org/}}.  The evolution of an
{\nbody} system is described by the second order ordinary differential equation

\begin{align}
    \bs{\ddot{r}}_{i} &= -G \sum\limits^{N}_{\substack{j=1\\j\neq i}}
                          m_{j} {(\bs{r}_i - \bs{r}_j)\over
                          | \bs{r}_i - \bs{r}_j|^{3}}\label{eq:nbody},
\end{align}

\noindent
where $G$ is the gravitational constant, $m_j$ is the mass of the $j$th particle
and $\bs{r}_j$ the position. We denote vectors with bold fonts.
The basis of the problem is purely dynamical, because the orbital evolution
is determined exclusive by the gravitational interaction.

The total energy of the system is a useful quantity to keep track of every
time step in the integration. It is given by the expression

\begin{align}
    E &= {1 \over 2} \sum\limits^{N}_{i=1} m_{i} \bs{v}_{i}^{2} -
         \sum\limits_{i=1}^{N} \sum\limits_{j > i}^{N}
         {G m_{i} m_{j} \over |\bs{r}_{i} - \bs{r}_{j}|},
\end{align}

\noindent
where $\bs{v}_i$ is the velocity of the particle $i$.

To numerically integrate the system of equations we adopt the 4th-order Hermite
integrator (H4 from now onwards) presented in \cite{Makino91,ma92} \citep[and see
also][]{Aarseth99,Aarseth03}.  H4 is a scheme based on a predictor-corrector
scenario, which means that we use an extrapolation of the equations of motion
to get a predicted position and velocity at some specific time. We then use
this information to get the new accelerations of the particles, later we
correct for the predicted values using interpolation based on finite
differences terms.  One can use polynomial adjustment in the gravitational
forces evolution among the time because the force acting over each particle
changes smoothly (which is the reason why adding a very massive particle
representing e.g. a super massive black hole will give you sometimes a
headache). {To advance the system to the following integration time
we approximate the equations of motion with an explicit polynomial.}
This prediction is less accurate, but it is improved in the corrector phase,
which consist of an implicit polynomial that will require good initial values
to scale to a good convergence.

This is a fourth-order algorithm in the sense that the predictor includes the
contributions of the third-order polynomial, and after deriving the
accelerations, adds a fourth-order corrector term. In the remaining of this
paper we focus on the implementation of the scheme into our GPU (and CPU) code
and how to maximise all of the computational resources available. For a detailed
description of the idea behind H4, we refer the reader to the article in which
it was presented for the first time, \citep{ma92}.

An advantage of the choice for H4 is that we can use the family of Aarseth's
codes (among others) as a
test-bed for our implementation.  These codes --some of which adopt H4, but not
all of them-- have been in development for more than 50 years. The codes are
public and have been widely used and validated, improved and checked a number
of times by different people, they have been compared to other codes and even
observational data. In this regard, to test our implementation and
parallelisation of H4, the access to the sources of the codes is an asset.

\subsection{Numerical strategy}

A main goal in the development of {\GR} is its {\it legibility}. We have focused in
making it easy to read and modify by other users or potential future developers
without compromising the computational performance of the algorithm.  This
means that we have made a significant effort in keeping a clear structure in
the source code so that, in principle, it can be well understood by somebody
who has not previously worked with it with relatively little effort.
The modularity of the code should allow new users to easily implement new
physics or features into it or adapt it to the purposes they seek.  It is
unfortunately easy --at least to a certain extent-- to miss either clarity in
coding or performance, when trying to have both in a code.  For instance, if we
want to obtain the best performance possible, one has to use low-level
instructions that for an outside user might result into something difficult to
understand when reading or trying to modify the source code. On the other hand,
name conventions for files, functions and variables might become a burden to
certain applications.

While most existing {\nbody} codes have achieved certain balance between the
two to some degree, it is difficult to adapt them to new architectures and
technology to boost their performance.
For the development of {\GR}, we have followed the next steps:

\begin{itemize}
    \item Serial Implementation of the initial version,
    \item Profiling and assessment of the integrator,
    \item Algorithm classification and finding the hot-spots,
    \item Optimisation of the bottlenecks.
\end{itemize}

\subsection{Particular choices}

\begin{description}

    \item[{\bf Object oriented programming:}]

Object oriented programming (OOP) is a powerful paradigm that allows us to
program an algorithm as objects interactions. In {\GR}, we use OOP.
The reason beneath it is related to our parallelisation scheme,
which is described below, more concretely with the data structure we have
chosen.

We have mainly two possible choices for data structures: classes with arrays, or
Arrays of Objects, which follows the basic idea of Struct of Arrays (SoA) and
Array of Structs (AoS).  For {\GR} we have chosen {\em classes} with arrays for
the main units of the program structure.  It is a good strategy to minimise the
data transfer between Host and Device, so as to avoid having large
communication times.

It is not required to update the forces of all the particles, so that we
encapsulate the information of the {\rm active} particles, and then we transfer
the AoS to the GPU.  All the remaining attributes of the bodies (i.e. those not
transferred to the GPU) are just class-members (arrays), and need to be in the
host CPU. An example of this could be large linear arrays, such as the time
steps of the particle.

    \item[{\bf Class distribution:}]

Since our code is using OOP, we describe a brief interaction between the
classes in Fig.~\ref{fig:classes}.  The main header, \texttt{common.hpp},
contains the definition of the constants, structures, macros, etc.  The idea
behind this model is to easily be able to add more features in upcoming
versions of our code, from new utilities functions to new integration schemes.

\begin{figure}
\resizebox{\hsize}{!}
          {\includegraphics[width=0.9\textwidth,clip]{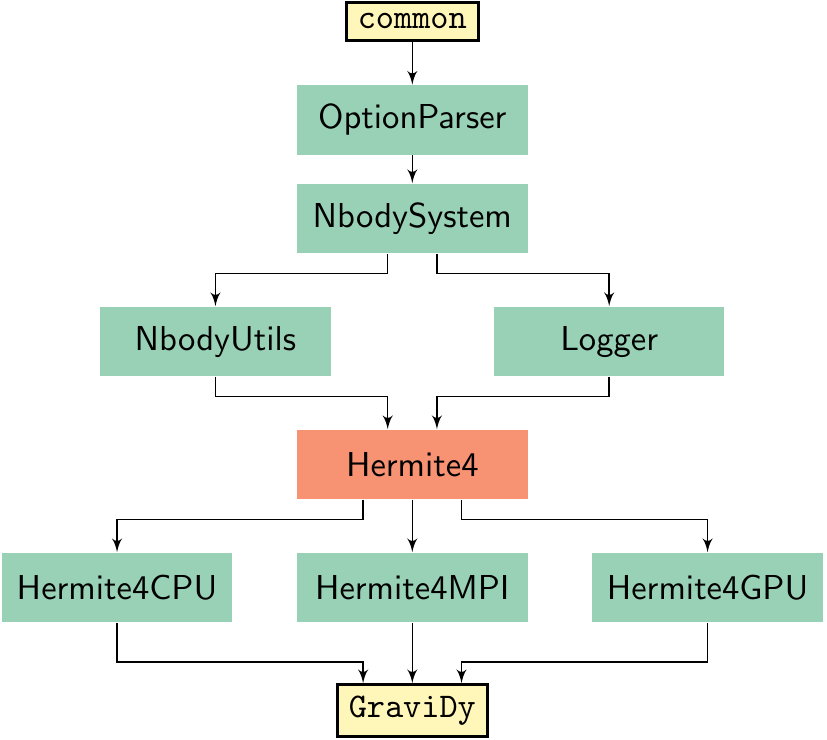}}
\caption{Class diagram of the code that shows the hierarchy of the application structure
({\GR}).}
\label{fig:classes}
\end{figure}

Every class is in charge of a different mechanism, from getting the integration
options from command-line, to the different integration methods using
parallelism or not~\footnote{For more information, please refer to the code
documentation}.

    \item[{\bf Double-precision (DP) over Single-precision (SP):}]

Using DP or SP in $N-$body codes has been already addressed by different
authors in the related literature~\citep[see e.g.][]{Hamada2007, keigo,
Gaburov2009}. Using DP is not the best scenario for GPU computing, because
there is a decrease factor in the maximum performance that a code can reach. We
can reach only half of the theoretical maximum performance peak, which depends
on each individual card: for example, the NVIDIA Tesla C2050/M2050 has a peak of
the processing power in GFLOPs $1030.46$ with SP, but only $515.2$ with DP.

{We choose DP for a more accurate numerical representation,
because it provides us a simple way of getting better energy conservation,
at the expenses of performance.}
There are different approaches, like the mixed-precision, \citep{Aarseth85},
and pseudo DP (\citealt{keigo}, currently used in the code $\phi-$GPU,
\citealt{berczik2011high}). These offer a relatively more accurate
representation (compared to SP) without a big impact in performance.

\end{description}


\subsection{The implementation scheme}
\label{sub:ImplScheme}

These are the steps that \GR~ follows when running a simulation:

\begin{enumerate}
    \item Memory allocation of the CPU and GPU arrays.
    \item Initialisation of the variables related to the integration.
    \item Copy the initial values of positions, velocities and masses
          of the particles to the GPU to calculate the initial system energy,
          and calculate the initial acceleration and its first time derivative,
          the so-called ``jerk''.
          The cost of this force calculation is $O(N^{2})$.
    \item Copy the initial forces from the GPU to CPU.
    \item Find the particles to move in the current integration time, $N_{\rm act}$,
          with a cost $O(N)$.
    \item Save the current values of the forces, to use them in the correction step,
          with a cost $O(N)$.
    \item Integration step:
          \begin{enumerate}
              \item Predict the particle's positions and velocity up to
                    the current integration time, with cost $O(N)$.
              \item Copy of the predicted positions and velocities of all the
                    particles from the CPU to the GPU.
              \item Update the $N_{\rm act}$ particles on the GPU,
                    which is explained in detail in section~\ref{sec:parallel}.
              \begin{enumerate}
                    \item Copy the $N_{\rm act}$ particles to a temporary array
                          on the GPU.
                    \item Calculate the forces between the particles on the GPU, with a cost
                          $O(N_{\rm act} \cdot N)$.
                    \item Reduce forces on the GPU.
                    \item Copy the new forces from the GPU to the CPU.
              \end{enumerate}
              \item Correct the position and velocity of the $N_{\rm act}$
                    updated particles on the CPU, $O(N_{\rm act})$.
              \item Copy the positions and velocities of the corrected $N_{\rm act}$
                    particles from the CPU to the GPU.
          \end{enumerate}
\end{enumerate}

{{\GR} adheres to the usual good practises of the beginning of the development
of every direct-summation $N-$body code:}

\begin{itemize}
    \item \emph{Direct-summation}, also known as particle-particle strategy,
          This approach is the simplest way to address
          the task of calculating the exerted force by all the $N-1$ bodies
          on a single body that we need to update at certain time step.
          This brute-force procedure has an order $O(N^{2})$,
          which represents the bottleneck of the algorithm.
    \item \emph{Softened point-mass potential}, {as an alternative in this version of the code
          to a proper close encounter regularisation.}
          All particles
          are represented by a dimensionless point mass.
          We introduce a softening parameter ($\epsilon$) in the
          distance calculation between two bodies while we get the
          new forces,

          \begin{align}
              \bs{\ddot{r}}_{i} &= -G \sum\limits^{N}_{\substack{j=1\\j\neq i}}
                                    {m_{j} \over (r^{2}_{ij} + \epsilon^{2})^{3/2} }
                                    \bs{r}_{ij},
          \end{align}

          \noindent
          so as to handle the situation in which two bodies get closer.

    \item \emph{Block time steps},
          It is not straightforward to have an {\nbody} code using individual
          time steps in parallel computing, because the idea behind massive
          parallelism is to perform the same task on different data chunks.
          We use the block time steps algorithm \citep{Press86}, to update group
          particles
          simultaneously. This scheme has been adopted by a number of authors
          \citep{portegies2001, hut2003,Aarseth99,Aarseth03,harfst2008,NitadoriAarseth2012}.

\end{itemize}

The algorithm uses a H4 to integrate the evolution.
The description of all the equations for each step is presented in
\cite{ma92}~\footnote{Eq. (2) has a typo in the sign of the second term in the
sum of $\bs{\dot{a}}_{i,1}$.}

\subsection{The parallelisation scheme}
\label{sec:parallel}

As we have already mentioned, the bottleneck of any {\nbody} code is the force
calculation.
In this respect, {\GR} is not different and a
quick performance test to get the profile of our serial code yields almost
$95\%$ of the execution time in this calculation.
We hence
introduce a parallelisation scheme, which we discuss in detail now.

{\GR} is based on a direct-summation H4 integrator and uses
block time steps, so that in the force update process we have a nested loop
for every $i-$active particle (which we will refer to from now with the
subscript ``act''). This means that for every particle which needs to be updated
we have a loop run on the whole set of particles of our system to check whether
$j-$particle is interacting with the $i-$particle.

\begin{figure}
\resizebox{\hsize}{!}
          {\includegraphics[width=0.9\textwidth,clip]{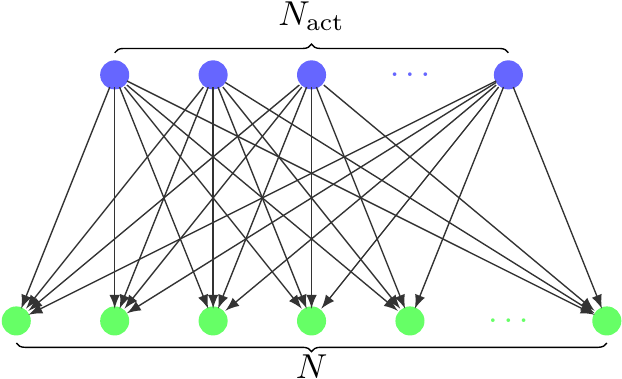}}
\caption
   {
Relation between the particles which will be updated in a certain
integration time ($N_{\rm act}$) and the whole set of particles ($N$).
The relation between the active particles and the others is
$N_{\rm act} \ll N$ in non-synchronisation times.
   }
\label{fig:nact_n}
\end{figure}

The whole process scales with the amount of $i-$particles,
as we can see in Figure~\ref{fig:nact_n}.

We then need to parallelise the loop corresponding to each of the
$i-$particles. For each of them we circulate through all of the $j-$particles,
and this is the process which needs to be parallelised.
Although this is in principle a straightforward scheme, since we
focus on GPUs, we run into the following issues:

\begin{enumerate}
    \item A GPU can launch a large number of threads, easily up to thousands of them.
          In our scenario, however, the number of active particles $N_{\rm act}$ is
          very small compared to the total amount of particles ($N$). This has an
          impact on the performance: we do not use all available threads, we are
          integrating a grid of $N_{\rm act} \times N $ forces. When the number of
          active particles is very low our \emph{occupancy} will be bad.

    \item Contrary, in the case in which we have to move all particles, we will have
          an $O(N^{2})$ parallelism, which maximises the GPU power. In this case, however,
          the \emph{memory bandwidth} is the limitation factor, since
          every particle requires all information about all other $N-1$
          particles.
\end{enumerate}

It is better to have all particles handled by the GPU, and not only the active
ones, because even though this subgroup is smaller, or even much smaller, it is
more efficient from the point of view of the GPU, since the \emph{occupancy} is
improved. The parallelisation happens at $j-$level (i.e. when calculating the
forces between active particles with the rest of the system). This idea was
first implemented by \cite{keigo}, and has proven to yield very good
performance.

\begin{figure}
\resizebox{\hsize}{!}
          {\includegraphics[width=0.9\textwidth,clip]{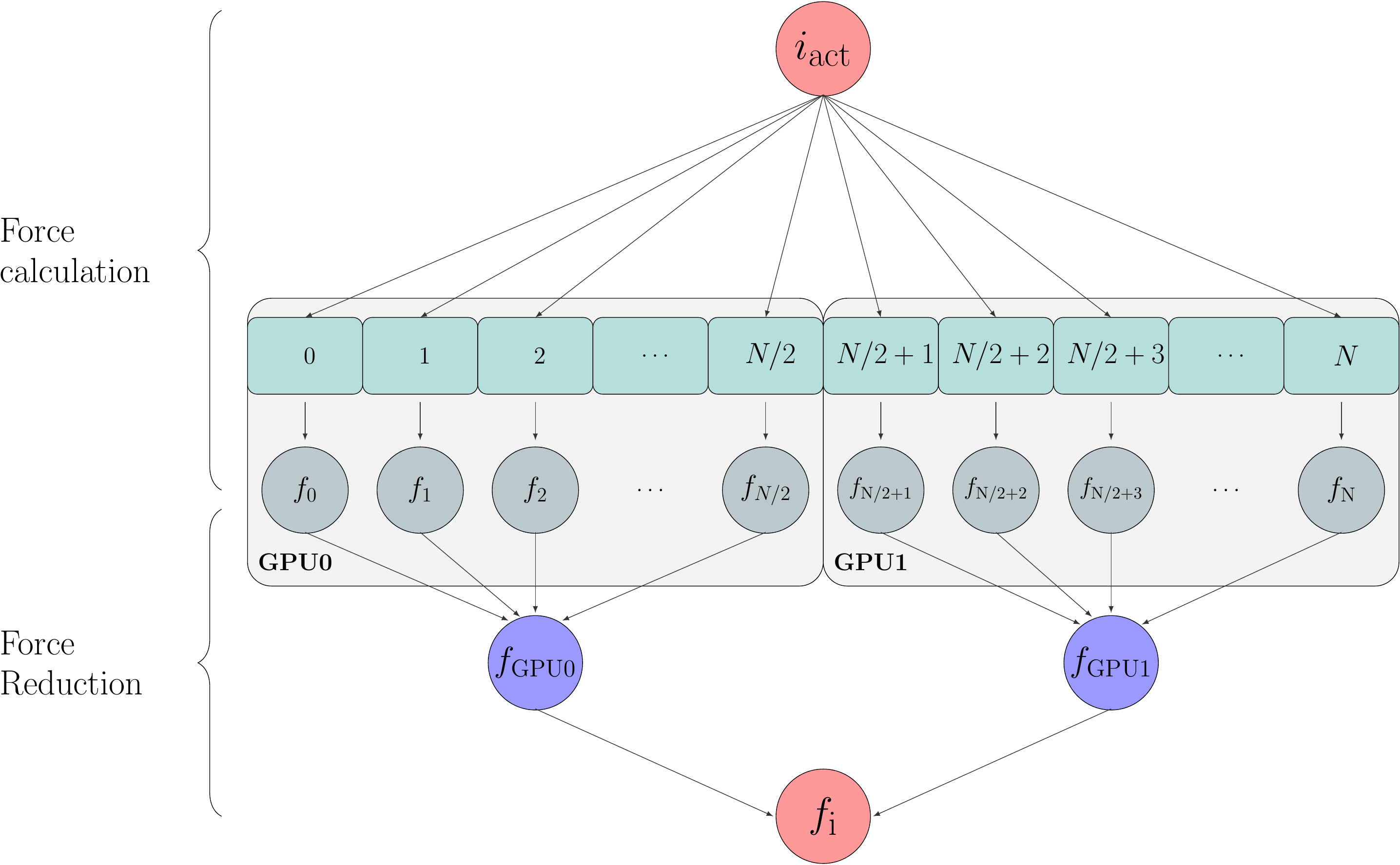}}
\caption
   {
parallelisation scheme to split the $j-$loop instead of the $i-$loop.
             Two GPUs are depicted to represent how the code works with
             multiple devices.
             In this case, we have two sections, the first is to calculate
             the force interactions of the $i-$particle with the whole
             system but by different threads (upper part).
             Then a reduction, per device, is necessary
             to get the new value for the $i-$particle force ($f_{i}$).
   }
\label{fig:force_split_reduction}
\end{figure}

The main ideas behind the $j-$parallelisation is how force calculation is done
and the summation of the forces (``reduction''):

\begin{itemize}
    \item \emph{Force calculation:} The interaction between the $i-$particle
        and the rest of the system is distributed among the GPU threads,
        which means that we launch $N$ threads, and each of them
        calculates its contribution with the $i-$particle.
        After this calculation, we have an array where each element
        contains the contributions all the particles.j
        This corresponds to the
        upper part of Fig.(\ref{fig:force_split_reduction}), which illustrates
        a set-up of two GPUs. After the force calculation we end up with an
        array containing the information about the forces for all particles.
    \item \emph{Force reduction:} In the lower part of the same Fig. we depict the
        summation of all of these forces, which is also performed in parallel,
        so that we use the blocks distribution of the GPU for this task.
\end{itemize}

\section{The three flavours of {\GR}: Tests}
\label{sec.experiments}

Thanks to the fact that there is a number of codes implementing similar
approaches to ours, we are in the position of running exhaustive tests on
{\GR}.  Indeed, the global dynamics of a dense stellar system (typically an
open cluster, because of the limitation in the number of particles we can
integrate) has been addressed numerically by a large number of authors in the
field of stellar dynamics. Therefore, we have decided to focus on the dynamical
evolution of a globular cluster with a single stellar population. We
present in this section a number of tests to measure the performance and the
accuracy of the three versions of {\GR} which we present using different amount
of particles. Our goal is to be able to offer an Open Source code that fits different needs
and requirements. This is why this first release of {\GR} offers three
different choices, which are general enough for different users with different
hardware configurations. These are:

\begin{description}

\item[\textbf{(i) The CPU version}] consists in the more basic implementation
in this work, a CPU version. I.e. This version uses
OpenMP and is intended for a system without graphic processing units, but with
many cores. This flavour can be used for debugging purposes by disabling the
OpenMP directives ({\tt \#pragma omp}). This is the basis for our further
development of the code.

\item[\textbf{(ii) The MPI version}] is virtually the same serial
implementation, but with OpenMPI directives added to improve the performance of
the hot-spots of the algorithm, in particular the force and energy calculation.
In this case we use the MPI library, and hence it can be run on a single
machine using a certain amount of cores as ``slave'' processes or on a large
cluster with separated machines as slaves.

\item[\textbf{(iii) The GPU version}] discards all CPU usage and only relies on
the GPU to integrate all gravitational interactions. As we mention later, we
tried to use CPU combined with GPU, but we did not see any benefit in it, and
the approach was hence neglected. We use CUDA to be able to interact
with NVIDIA graphics processing units. The code is designed
to detect the amount of present GPUs and use all of them, unless otherwise
required by the user. This means that this version can use in a parallel way as
many GPU cards as the host computer can harbour in a very simple and efficient
way. The communication
between the different GPU cards in the host computer is internal and run
through Peripheral Component Interconnect Express (PCIe), a high-speed serial
computer expansion bus standard, so that the data flows rapidly because of the
low overhead.

\end{description}

The specifications of the hardware (CPU, GPU and available RAM) and operating
systems we used are summarised in table~\ref{tab:hw}.

\begin{table}
    \Huge
    \centering
    \resizebox{\hsize}{!}{
    \begin{tabular}{lr}
        \hline
        {\bf System A} & \texttt{datura} (165 nodes) \\
        \hline
        {CPU} &  Intel(R) Xeon(R) CPU X5650 @ 2.67GHz (24 cores)\\
        {GPU} &  \emph{none}\\
        {RAM} &  24 GB \\
        {OS}  &  Scientific Linux 6.0 \\
        \hline
        {\bf System B} & \texttt{gpu-01} (1 node)\\
        \hline
        {CPU} &  Intel(R) Xeon(R) CPU  E5504  @ 2.00GHz (4 cores)\\
        {GPU} & 4 x Tesla M2050 @ 575 Mhz (448 cores) \\
        {RAM} &  24 GB \\
        {OS}  &  Scientific Linux 6.0 \\
        \hline
        {\bf System C} & \texttt{krakatoa} (1 node)\\
        \hline
        {CPU} & AMD Opteron 6386SE @ 2.8 GHz (32 cores) \\
        {GPU} & 2 x Tesla K20c @ 706 MHz (2496 cores)\\
        {RAM} & 256 GB \\
        {OS}  & Debian GNU/Linux 8 \\ \\
        \hline
        {\bf System D} & \texttt{sthelens} (1 node) \\
        \hline
        {CPU} & Intel(R) Xeon(R) CPU E5-2697v2 (IvyBridge) @ 2.7GHz (24 cores) \\
        {GPU} & 2 x Tesla C2050 / C2070 @ 1.15 Ghz (448 cores)  \\
        {RAM} & 256 GB \\
        {OS}  & Debian GNU/Linux 8 \\
        \hline
    \end{tabular}}
    \caption{Specification of the different systems of the Albert Einstein Institute
             used for the tests.
             }
    \label{tab:hw}
\end{table}

\subsection{Initial conditions and {\nbody} units}

For all of our tests we choose an equal-mass Plummer sphere~\citep{plummer1911}
for the sake of comparison with other codes.
We choose standard {\nbody}
units (NBU, hereon) for the calculations and in the resulting
output~\citep{Henon,Heggie1986}. This means that

\begin{itemize}
    \item The total mass of the system is 1: $\sum^{N}_{\rm i=0} m_{i} = 1$.
    \item The gravitational constant (G) is set to 1: $G=1$.
    \item The total energy of the system is equal to $-0.25$:
          $E_{\rm tot} = K + U = -0.25$, with $K$ and $U$ the total
          kinetic and potential energy of the system, respectively.
    \item {The virial radius is set to $\approx 1$.}
\end{itemize}

{The Plummer spheres have a fixed half-mass radius of $0.8$ and a Plummer
radius of $0.6$.}

We used the code by \cite{mcluster} (McLuster) to generate all the initial
conditions for the test we performed on the current work.

\subsection{Accuracy, performance and speed}

For {\GR}, as we have seen, we have chosen a H4 integrator.  The
numerical error introduced scales hence as $O(\Delta t^{4})$ assuming a shared
time step, which means that the previous is true {\em only} if all particles are
updated at every integration step.  Since we use a block time step scheme,
certain groups of particles share a time step value, but not all of them.
Thanks to this approach, the numerical error which we reach in our integrations
is slightly less than the value mentioned previously.

\begin{figure}
\resizebox{\hsize}{!}
{\includegraphics[width=1.1\textwidth,clip]{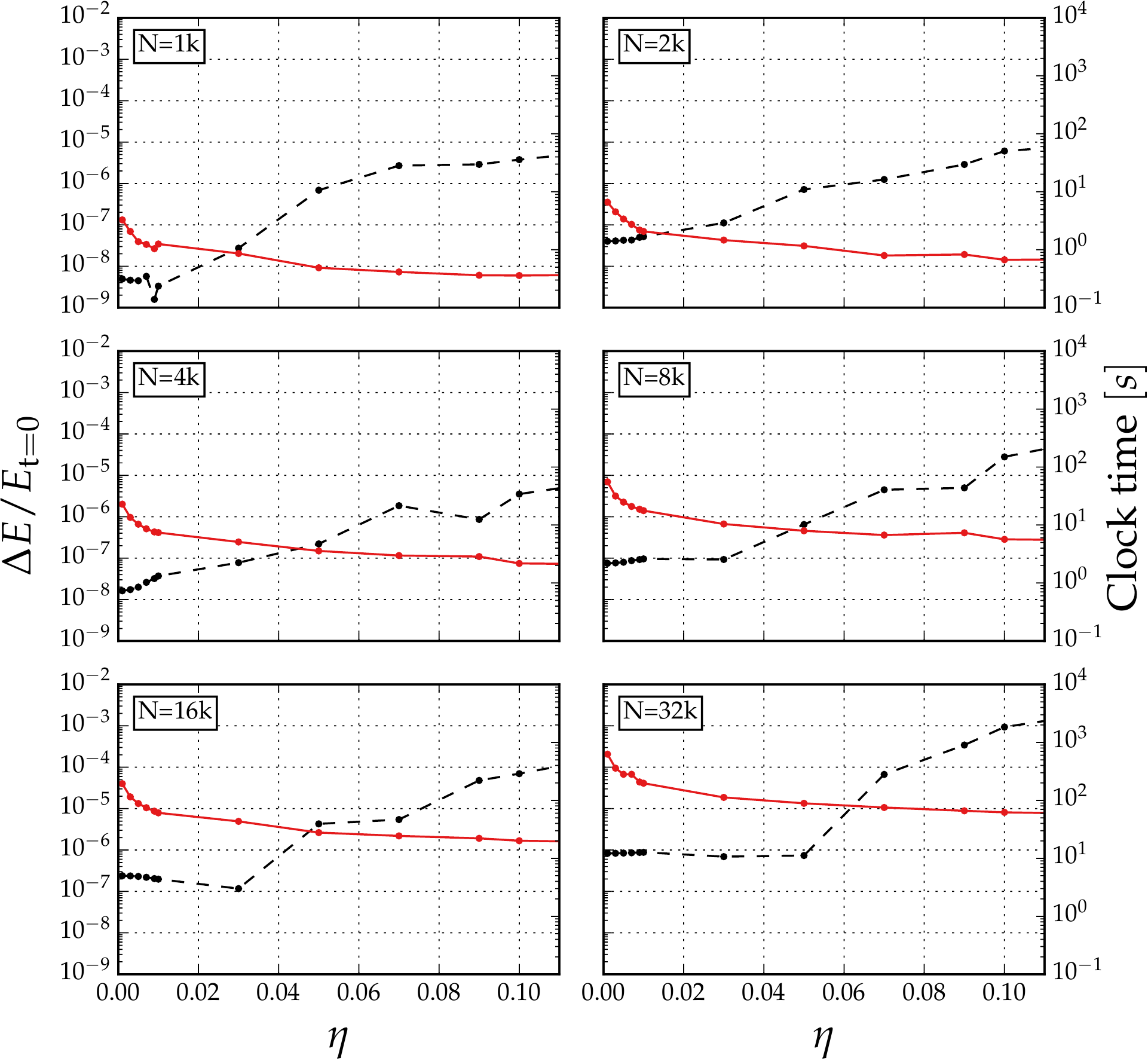}}
\caption
    {
Cumulative energy error (dashed black line) and wall clock time (solid red line)
in function of $\eta$ for six different systems consisting of a Plummer sphere
with $N=1,\,2,\,4,\,8,\,16,\,32$\,k particles, with $k:=1000$, from the top to the bottom, left
to right. The integration corresponds to one time unit, namely from $t=1$ to
$t=2$ in the wall clock time analysis, and for $t=2$ in the energy error calculation.
The reason for choosing the elapse between 1 and 2 is to get rid of any initial
numerical error at the simulation startup, from 0 to 1. All tests have been
performed on System B of Tab.~(\ref{tab:hw}).
    }
\label{fig:eta_energy_time}
\end{figure}

A free parameter, $\eta$, was introduced in ~\cite{ma92}
responsible for determining the calculation of every time step of the system,
from the initial calculation to the update after every iteration. Hence, so as
to assess an optimal value for it, we perform different tests to find a balance
between a good energy conservation and a minimum wall clock time.  We
explore values between $0.001$ and $0.1$ integrating a series of systems with
$N$ ranging between 1024 to 32768,
for convenience~\footnote{Any number of particles can be also handle properly,
not necessarily powers of 2.}, and up to 2 NBU. We show the results
in Fig.(\ref{fig:eta_energy_time}) performed on System B of Tab.~(\ref{tab:hw}). For
small values of $\eta$, the cumulative energy error approximately stays
constant, because the error is small enough to leave accuracy in hands of the
integrator scheme and the hardware.  Increasing $\eta$ leads to larger errors.
This is particularly evident when we use systems with a larger number of
particles. The system with $N=32768$ particles, and a $\epsilon = 10^{-4}$,
achieves $\Delta E/E_{0} \approx
10^{-3}$ for $\eta = 0.1$, while it is as low as $\Delta E/E_{0} \approx
10^{-6}$ for the same value and 1024 particles.

In the same figure we describe the performance in function of $\eta$ by using
the wall clock time in seconds for the code to reach one NBU for the same values of
the parameter. We can see that the value of $\eta$ is inversely proportional
to the time, since increasing its value results in decreasing the execution
time. When we increase $\eta$ we implicitly increase the time step of every
particle, so that one unit of time is reached sooner. We find that a value of
about $0.01$ is the best compromise for most of our purposes, yielding an
accuracy of about $\Delta E/E_{0} = 10^{-7}$ in most of the cases.

\begin{figure}
    \centering
    \includegraphics[width=0.4\textwidth]{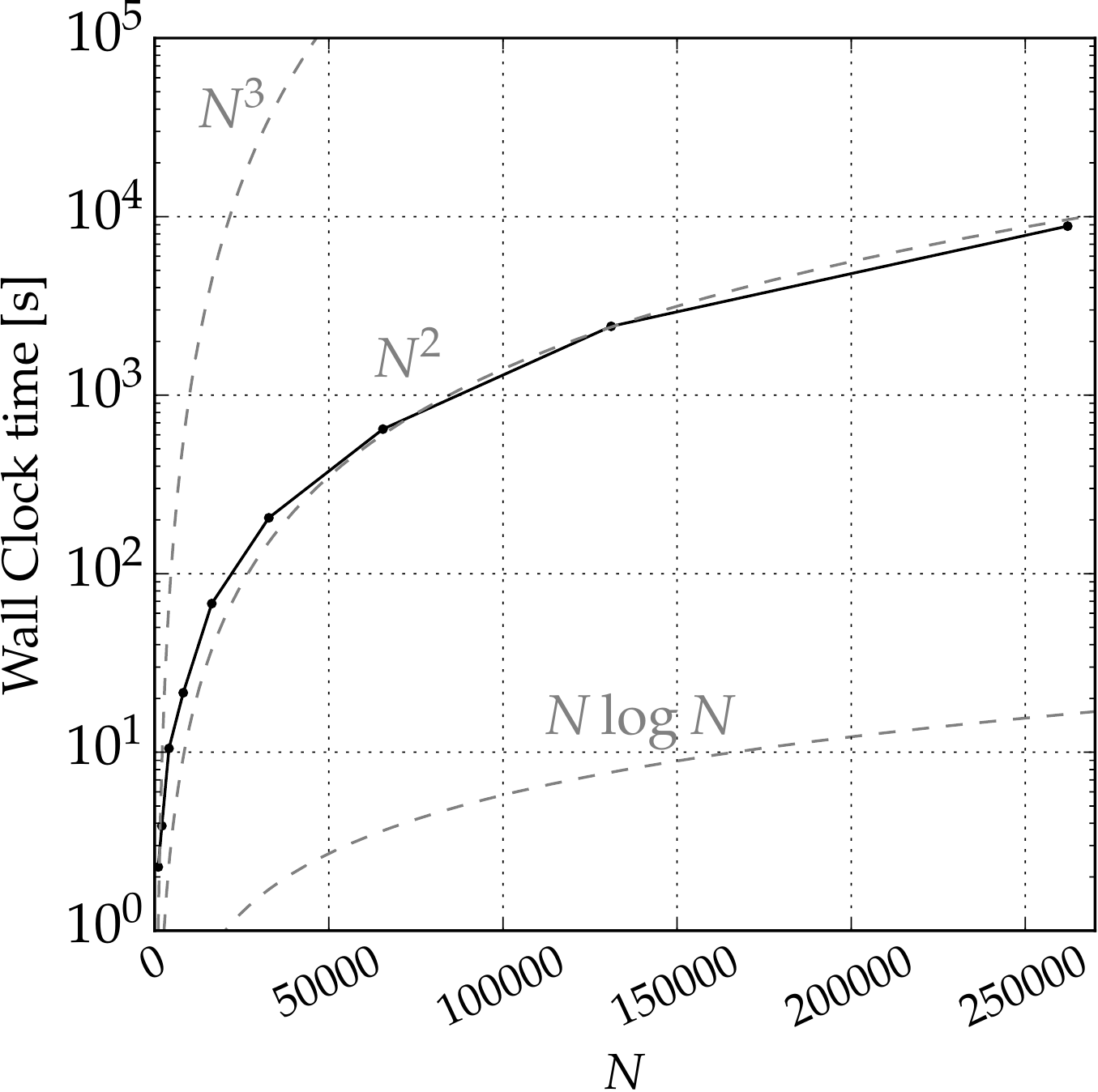}
    \caption{
Wall clock time of integration from $t=1$ NBU up to $t=2$ NBU, using $\eta = 0.01$
and $\epsilon = 10^{-4}$ using different amount of particles on System C of
Tab.(\ref{tab:hw}).
}
    \label{fig:time}
\end{figure}

To measure the execution speed of our code we perform a set of tests by
integrating the evolution for one NBU of a Plummer sphere with different
particle numbers, ranging from $N=1024$ to $N=262144$. For the analysis, we
choose the time starting at $t=2$ and finishing at $t=3$, since the first time
unit is not representative because the system can have some spurious numerical
behaviour resulting from the fact that it is not {\it slightly} relaxed.
When testing the parameters $\eta$ and $\epsilon$, we picked the time starting
at $t=1$ and finishing at $t=2$
because we wanted to understand their impact not right at the beginning of the
simulation. Now we allow the system to further relax so as to obtain a more
realistic system. In particular,
the distribution time steps drifts away from the initial setup.

We display the wall clock time of each integration in Fig.~(\ref{fig:time}). We also
display reference curves for the powers of $N^{3}$, $N^{2}$ and $N\log N$,
multiplied by different factors to adapt them to the figure. We see that {\GR}
scales very closely as a power of 2. The deviations arise from the fact that not
all particles are being updated at every time step.

\begin{figure*}
\resizebox{\hsize}{!}{
    \includegraphics[width=0.98\textwidth]{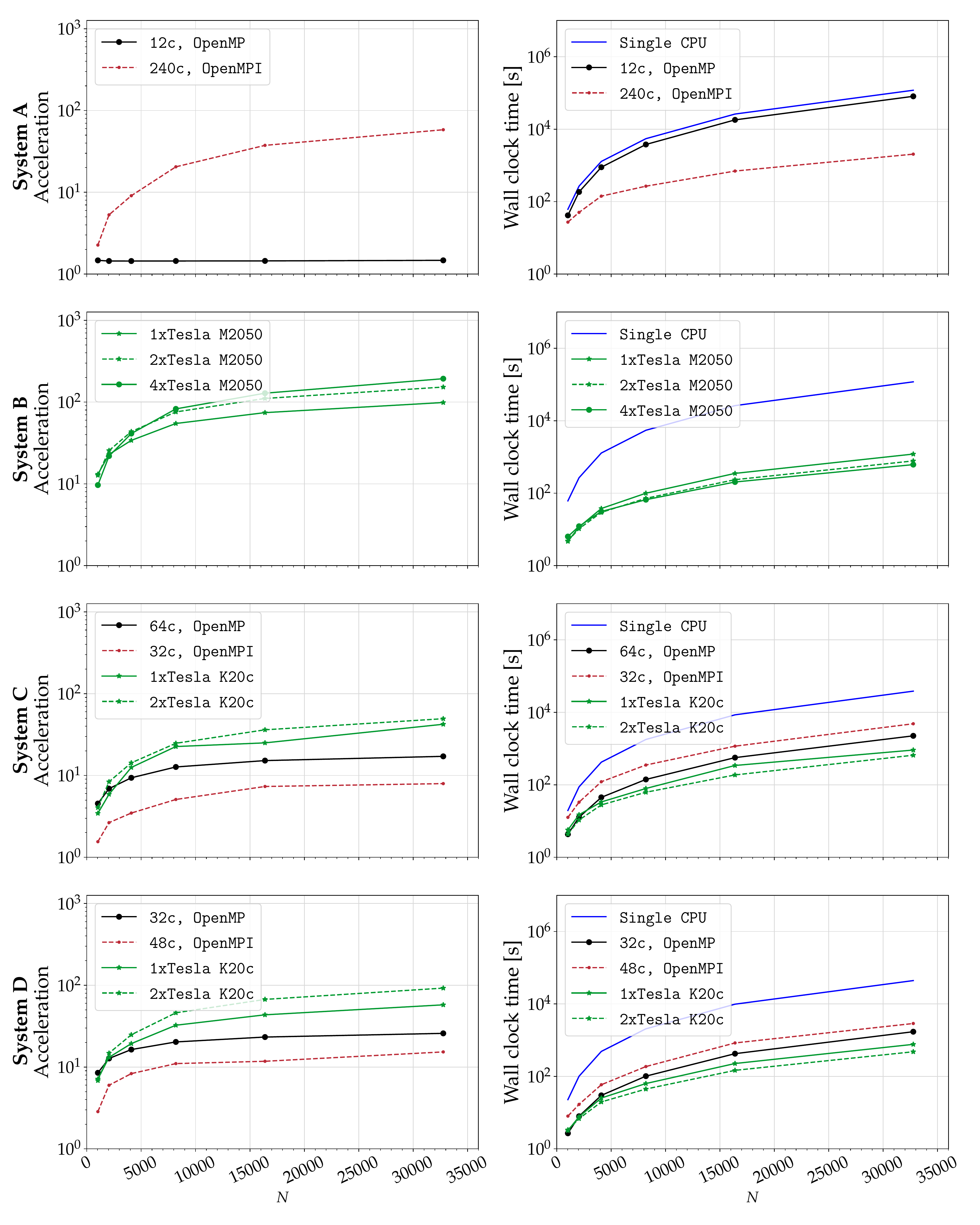}}
    \caption{Acceleration factor and wall clock time for the different parallel versions of the integrator
    (see Tab. \ref{tab:hw}). The acceleration factor is normalised to the single CPU version (1 thread), up to $T = 3 [NBU]$.
    For the CPU parallel version of the code, we give information about the number of cores with
    the letter ``c''. The GPU-parallel cases display the information on the number of cards with
    multiplying numbers.
    }
    \label{fig:accall}
\end{figure*}

In Figure~\ref{fig:accall} we show the acceleration factor for all parallel
scenarios as compared to the single-thread CPU case, which we use as a
reference point. Due to the design of the code, the maximum performance is
achieved with the larger particle number. The most favourable scenario for {\GR}
is, as we can see in the figure, System B. The 4 GPUs available boost the
performance up to a factor of 193 as compared with the single-thread CPU case.
A similar speed up is achieved on System D, which reaches a factor of 92 for
the 2 GPUs. The CPU-parallel version lies behind this performance: only
reaching a factor of 58 for System A, using up to 240 cores.

\subsection{Scaling of the three different flavours of the code}

An obvious question to any user of a numerical tool is that of scaling. In this
subsection we present our results for the three different versions of {\GR} of
how wall clock time scales as a function of threads or cores, or what is the
acceleration of the multiple-GPU version of the code in function of the
particle number as compared with a single GPU run, which we use as reference
point.

\begin{figure*}
\resizebox{\hsize}{!}
          {\includegraphics[width=0.9\textwidth,clip]{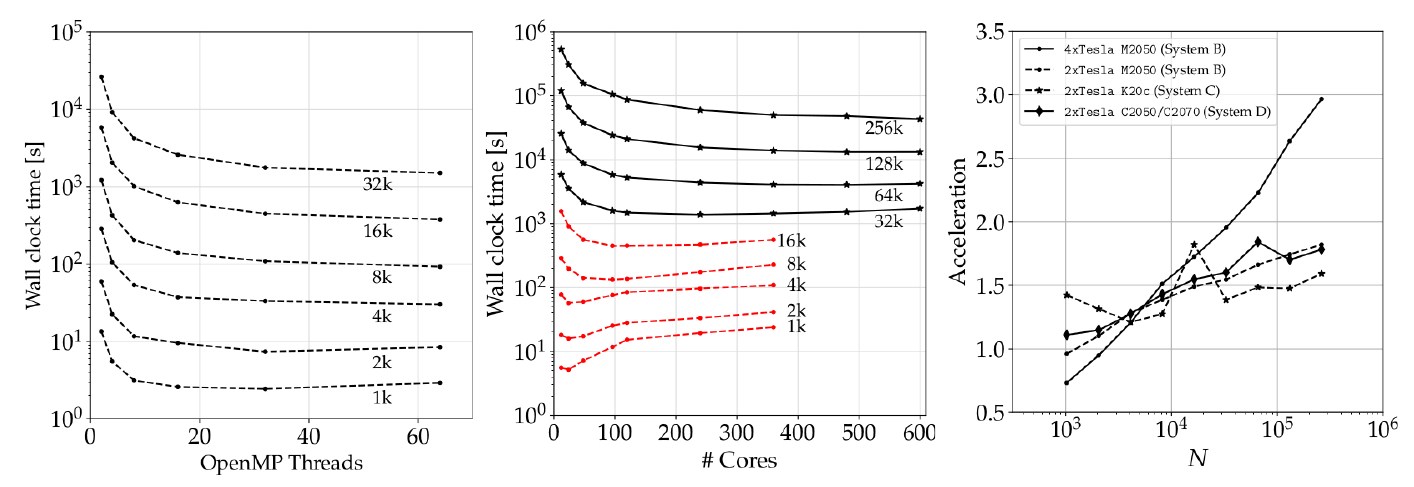}}
\caption
   {
Performance of the three different flavours of {\GR}, as a function of the number
of OpenMP threads, and number of cores and GPUs, for the CPU, MPI and GPU versions,
respectively and from left to right.
The integration corresponds to up to $t=2$ NBU.
{\em Left panel:} The CPU version runs on a single node with different numbers of
threads. The experiments were performed on system C of Tab.(\ref{tab:hw}).
{\em Mid panel:} The MPI version running on different numbers of
cores, using up to 600 of them and particles, up to 262144.
In this case we use system A of the same table.
{\em Right panel:} The GPU flavour using different amount of devices in parallel
and particles.
We show the acceleration factor as compared to a single GPU run for three
different setups with different GPU specifications, corresponding to systems
B, C and D of the same table.
   }
\label{fig.scalings}
\end{figure*}

In Fig.~(\ref{fig.scalings}) we depict this information for the CPU, MPI and
GPU versions. We can see in the CPU version that for small amounts of
particles, in particular for 2k and 1k, we have an increase in the execution
time with more threads, contrary to what we would expect. This is so because
the amount of parallelism is not enough and the code spends more time splitting
data and synchronising threads than performing the task itself, a usual
situation in tasks with a low level of computation.

The MPI version uses the same j-parallelisation idea from the GPU one. In this
case the code splits the whole system to the amount of available slaves (be it
cores or nodes), performs the force calculation and finally sums up
(``reduces'') the final forces for all active particles.
This procedure was performed developing our own
forces datatype operations and reduction, based on structures.
This means that we define our own operations to be able to ``sum'' two forces
(which are two three-dimensional arrays per particle).
The simulations with small amount of
particles (1k, 2k, 4k, 8k and 16k) are a clear example of a parallelisation
``overkill'': using more resources than what is actually needed. Additionally,
the communication process plays a role in scaling, which can be seen in the
curves corresponding to these simulations for a number larger than 200 cores -
the execution time increases instead of decreasing.  On the other hand, large
amount of particles (cases with 32k, 64k, 128k and 256k) show the expected
behaviour, a better execution time with more nodes or cores. Surely this is
not a solution for all simulations, since at some point the curves flatten.

The GPU version is a different scenario, since every device has its own
capability, limitations and features that makes it difficult to compare their
performances. For this reason we have decided to present the acceleration
factor of every case normalised to a single-GPU run in the same system.  This
flavour of {\GR} should always have a better performance when increasing the
particle number. Although having a good occupancy is in principle the ideal
scenario in this case, it is not necessarily the best reference point to assess
the efficiency of the CUDA kernels, because it is related to register
uses, but also to the amount of redundant calculations and the arithmetic
intensity.  We show the acceleration factor of two and four {\tt Tesla M2050}
devices as compared to a single-GPU run which have hardware and performance
differences~\footnote{The primary difference is that model M is designed for Original
Equipment Manufacturer (OEM) for an integrated system, without active cooling,
while model C includes the active cooling and can be installed on any standard
computer.} but they nonetheless reach a similar acceleration factor.
{We have access to two {\tt Tesla K20c},
which have more than the double peak performance in double precision floating point
compared to the other mentioned models. The scaling between using one
and two devices has a factor of 1.6.}

{
Every GPU is a different device, so that in order to obtain a proper optimisation
we need to first do a study in terms of kernel calls configuration.
The current work present a fixed configuration of 32 threads per block,
using a number of blocks corresponding to $N/32$.
A deeper study on each GPU-device configuration is planned for future
publication, where speeding up the first GPU implementation will be one of the
main concerns.}


\section{The role of softening on dynamics}

For the current version of {\GR}, and quoting Sverre Aarseth on a comment he
got some years ago during a talk, ``we have denied ourselves the pleasure of
regularisation''\citep{KS65,AarsethZare74,Aarseth99,Aarseth03}. This means that
the code resorts to softening, via the parameter $\epsilon$.
This quantity can be envisaged as a critical distance
within which gravity is, for all matters, nonexistent. This obviously solves
the problem of running into large numerical errors when the distance between
two particles in the simulation become smaller and smaller, because since they
are 0-dimensional, this induces an error which grows larger and larger as they
approach. This comes at a price, however. The relaxation time of the system
is, approximately \citep[see e.g. section on Two-body relaxation
by][]{Amaro-SeoaneLRR2012},

\begin{equation}
t_{\rm rlx} \sim N_* \, \frac{t_{\rm dyn}}{\ln(d_{\rm max}/d_{\rm min})}.
\end{equation}

In this equation  $d_{\rm min}$ and $d_{\rm max}$ are the minimum and maximum
impact parameters. In an unsoftened $N-$body problem they are of the order of
$d_{\rm min} \approx G\,m/\sigma^2$, and the size of the cluster, respectively.
In other words, $d_{\rm min} \approxeq R_{\rm cl}/N$, with $R_{\rm cl}$ the
radius of the self-gravitating cluster, if the system is virialised, and
$d_{\rm max}$ is of the the half-mass radius order.  Now suppose the code
uses a softening parameter $\epsilon$. If the value of $\epsilon$ is smaller
than $d_{\rm min}$, then softening should play only a minor role in two-body
relaxation, and the global dynamical evolution of the cluster must be similar
to that of another cluster using regularisation. In the contrary case in which
$\epsilon > d_{\rm min}$, the relaxation time is artificially modified, as we
can read from the last equation. The larger the quantity $\ln(d_{\rm
max}/d_{\rm min})$, the more efficient is relaxation, and hence the shorter the
relaxation time.

\subsection{``Best'' value for the softening?}

\begin{figure}
\resizebox{\hsize}{!}
{
\includegraphics[width=1.1\textwidth,clip]{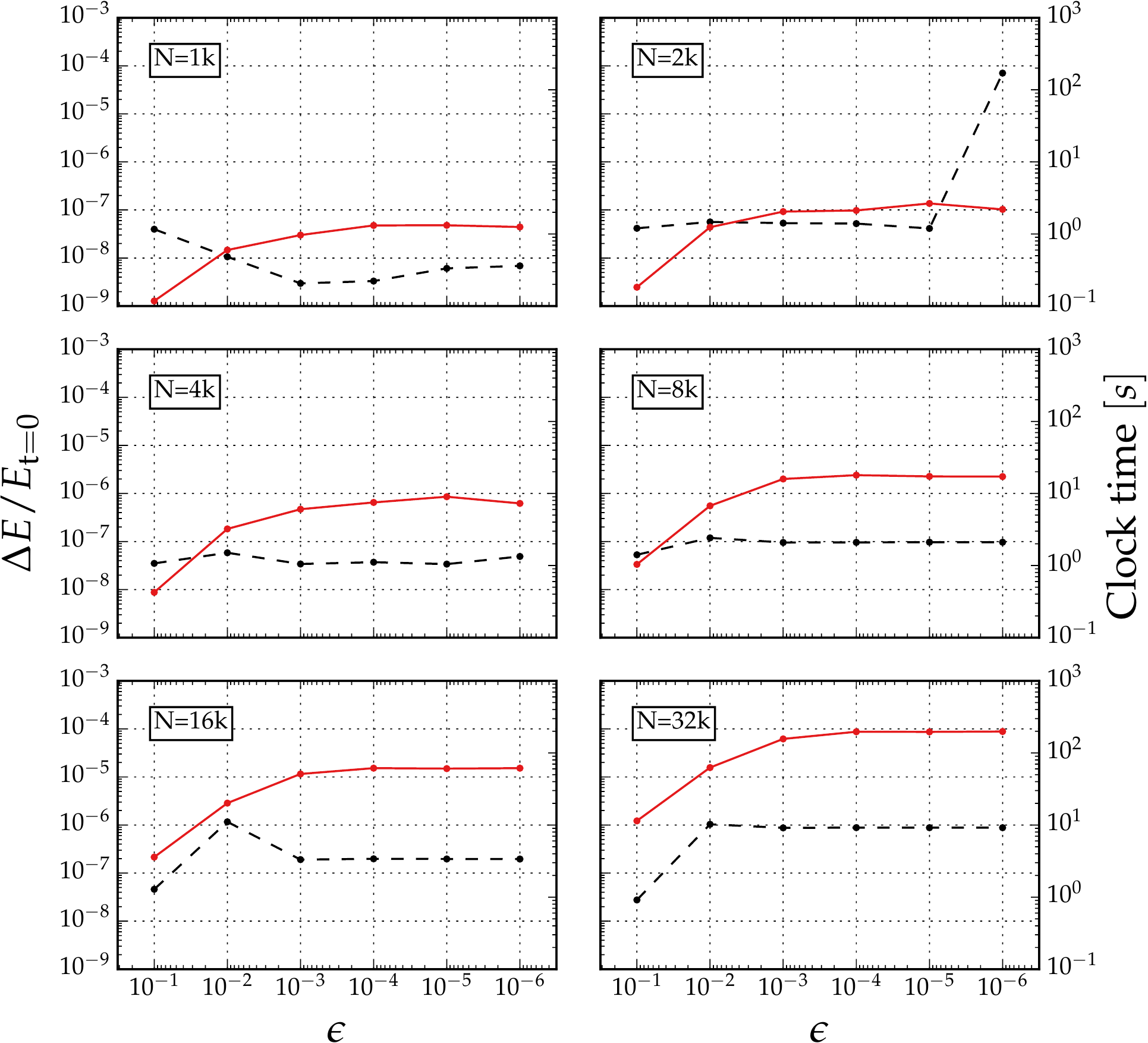}}
\caption
{
Cumulative energy error (dashed black line) and wall clock time (solid red line)
using different values of the softening ($\epsilon$).
We integrate different amounts of particles $N$ up to $t=2$ NBU.
The wall clock time corresponds to the execution time between $t=1$ and $t=2$ NBU
while the energy error is the one at $t=2$ NBU.
}
\label{fig:soft_energy_time}
\end{figure}

We perform a series of simulations to assess the relevance of $\epsilon$ in the
global dynamical evolution of an autogravitating stellar system.
In Figure~\ref{fig:soft_energy_time} we depict the energy error and wall clock time
for six different particle numbers as a function of the softening. The lower
its value, the faster the simulation.
{However, by using larger
values of the softening, we must understand that we are evolving a system in
which two-body deflections are not being taking into account.
This is the most important aspect of two-body relaxation, and therefore
a critical factor in the general evolution}.
Thus, the fundamental feature which drives
the global evolution of the system is non-existing below larger and larger
distances. In particular, the larger values correspond to about 10\% of the
virial radius of the system.
From
these panels it seems that a value of $\epsilon \approx 10^{-4}$ is a good
compromise \textit{for this particular test that we are running in this
example}. A good practice would be that the user tests different softening
values for the case which is being addressed before making a decision for the
softening. This choice is left for the user of the code, because we deem it
difficult, if not impossible, to implement a self-regulating scheme
in which the best value for the softening is calculated a priori.

\subsection{Core collapse}

\subsubsection{Single-mass calculations}

\begin{figure}
\centering
\resizebox{\hsize}{!}
          {\includegraphics[width=0.99\textwidth,clip]{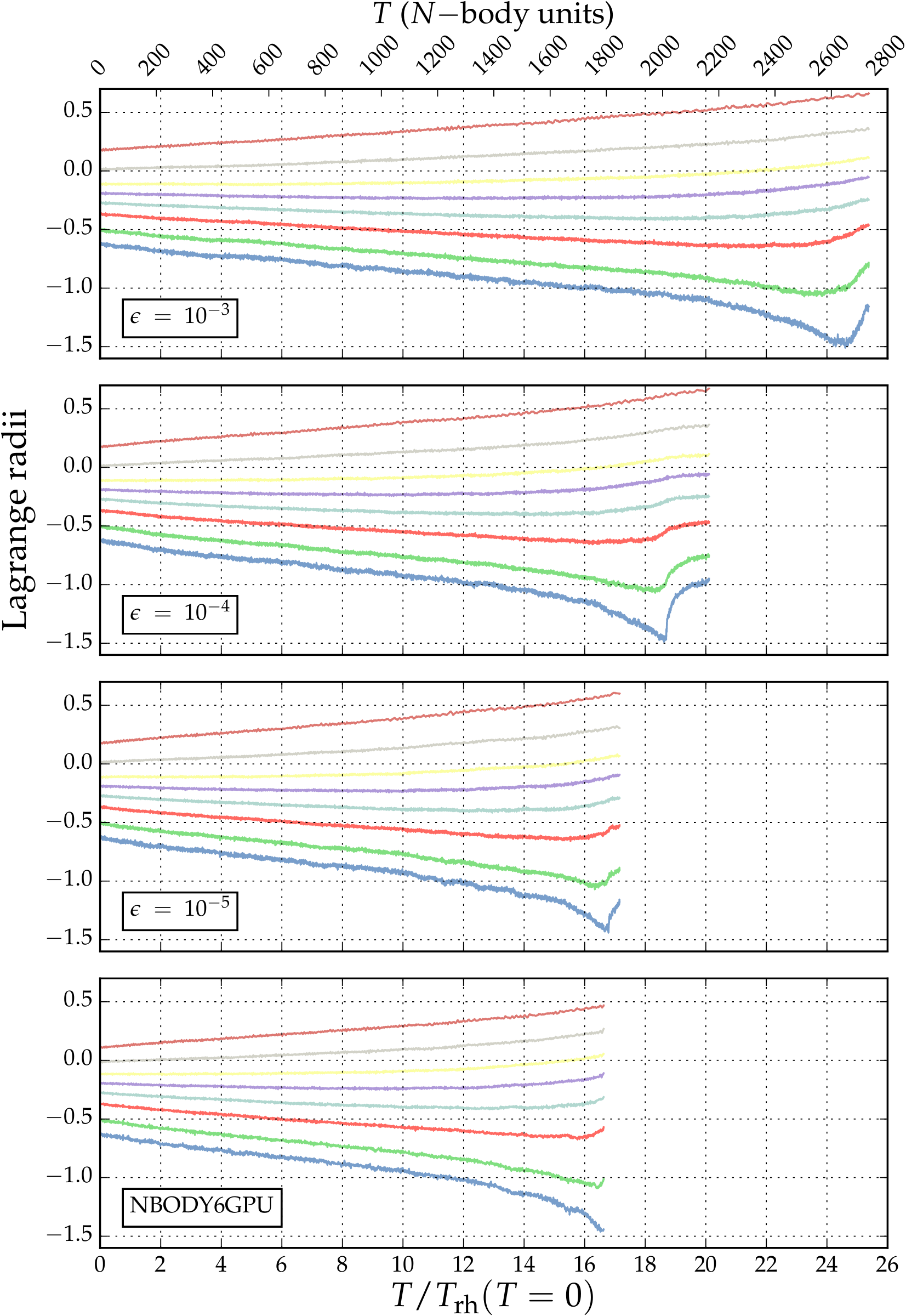}}
\caption
{
Comparison of the Lagrange radii of a Plummer Sphere with $N=8192$ particles,
using different values of $\epsilon$ (softening) for {\GR} and the {\nbg} code,
from upper to bottom.
The mass percentages are $0.5,\, 1,\, 2,\, 3,\, 4,\, 5,\, 6,\, 75$ and
$90\,\%$ of the total mass, from the bottom to the upper part of each plot.
The core collapse is reached at $\approx 24, 18 $ and $16\ T_{\rm rh}$ for
$\epsilon = 10^{-3}, 10^{-4}$ and $10^{-5}$ respectively.
The half-mass relaxation time for this system is $T_{\rm rh} = 112.186 [NBU]$
The {\nbg} code does not include a softening parameter, and treat binary
evolution with a KS-regularisation.
}
\label{fig:lagrangeradii}
\end{figure}

A good reference point to assess the global dynamical evolution of a dense
stellar system is the core collapse of the system \citep[see
e.g.][]{Spitzer87,Aarseth74,Giersz94}. We present here the evolution of the
so-called ``Lagrange radii'' (the radii of spheres containing a certain mass
fraction of the system) in Figure~\ref{fig:lagrangeradii}, for three
representative values of the softening, the three upper panels, as calculated
with {\GR}, and depict also the results of one calculation performed with
{\nbg} \citep{NitadoriAarseth2012}, the lower panel, which uses KS
regularisation \citep{KS65,Aarseth03}. This can be envisaged as the ``best
answer'', which provides the reference point with which the other calculations
should be compared.
In the figures we use the half-mass relaxation time, which we introduce as

\begin{align}
    t_{\text{rh}} &= 0.138 \left( { N r_{h}^{3}\over Gm } \right)^{{1\over 2}}
                      {1 \over \ln ( \Lambda ) },
\end{align}
\noindent
where $N$ is the number of particles of the system, $m$ the average mass of a star,
$r_{h}$ the half-mass radius,
and  $\Lambda := \gamma N$, with $\gamma = 0.1$
the argument of the Coulomb logarithm.

From the panels we can easily see the impact of the softening parameter in the
calculations: the collapse of the core is retarded for larger values. Our
default choice for the softening, $10^{-4}$ is just $2\ T_{\rm rh}$ earlier
than a {\nbg} calculation { that we performed to compare with our code}.

\begin{figure}
\centering
\resizebox{\hsize}{!}
          {\includegraphics[width=0.9\textwidth,clip]{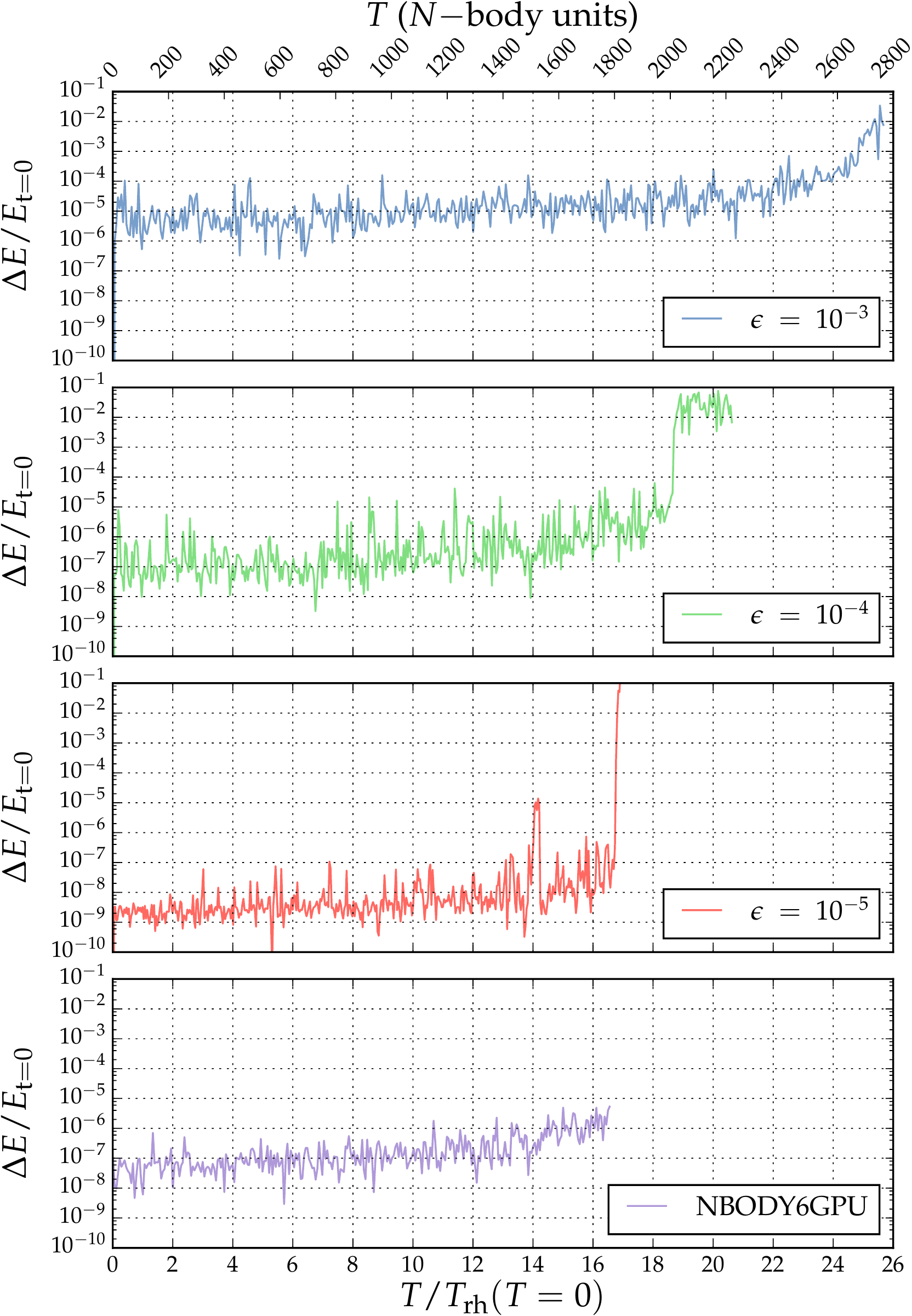}}
\caption
   {
Energy conservation in a long time integration of a system with $N=8102$
Comparison of the Energy conservation of a Plummer Sphere with $N=8192$ particles,
using different values of $\epsilon$ (softening) for {\GR} and the {\nbg} code,
from upper to bottom.
The core collapse is reached at $\approx 24, 18 $ and $16\ T_{\rm rh}$ for
$\epsilon = 10^{-3}, 10^{-4}$ and $10^{-5}$ respectively.
The half-mass relaxation time for this system is $T_{\rm rh} = 112.186 [NBU]$
The {\nbg} code does not include a softening parameter, and treat binary
evolution with a KS-regularisation.
All the runs were stopped after the core collapse.
   }
\label{fig:energy}
\end{figure}

Another way of looking at the core collapse is in terms of energy. In
Figure~\ref{fig:energy} we display the evolution of the energy for the same
systems of Figure~\ref{fig:lagrangeradii}.  As the collapse develops, the
average distance between particles becomes smaller and smaller.  There is an
obvious correlation between the conservation of energy and the value of the
softening.  The transition between a fairly good energy conservation and a bad
one happens more smoothly for larger and larger values of the softening, since
the error has been distributed since the beginning of the integration. This
means that, the smaller the value of the softening, the more abrupt the
transition between the good and bad energy conservation, which leads to a big
jump for the lowest value, $10^{-5}$. We stop the simulations at this point
because of the impossibility of {\GR} to form binaries, the main way to stop
the core collapse.

As discussed previously, and as we can see in  Figures~(\ref{fig:energy},
\ref{fig:lagrangeradii}), the introduction of softening in the calculations has
an impact on the global dynamical behaviour of the system.  We find factors of
1.001, 1.08 and 1.55 of delay to reach the core collapse for the softening
values  $\epsilon=10^{-5}$, $\epsilon = 10^{-4}$ and $\epsilon = 10^{-3}$,
respectively.

The {\nbg} simulation was run on a different system, using a {\tt
GeForce GTX 750 (Tesla M10)} GPU, which is why we compared
with the overall system evolution instead of the wall clock time.

\subsubsection{Calculations with a spectrum of masses}

\begin{figure*}
\resizebox{\hsize}{!}
          {\includegraphics[width=0.9\textwidth,clip]{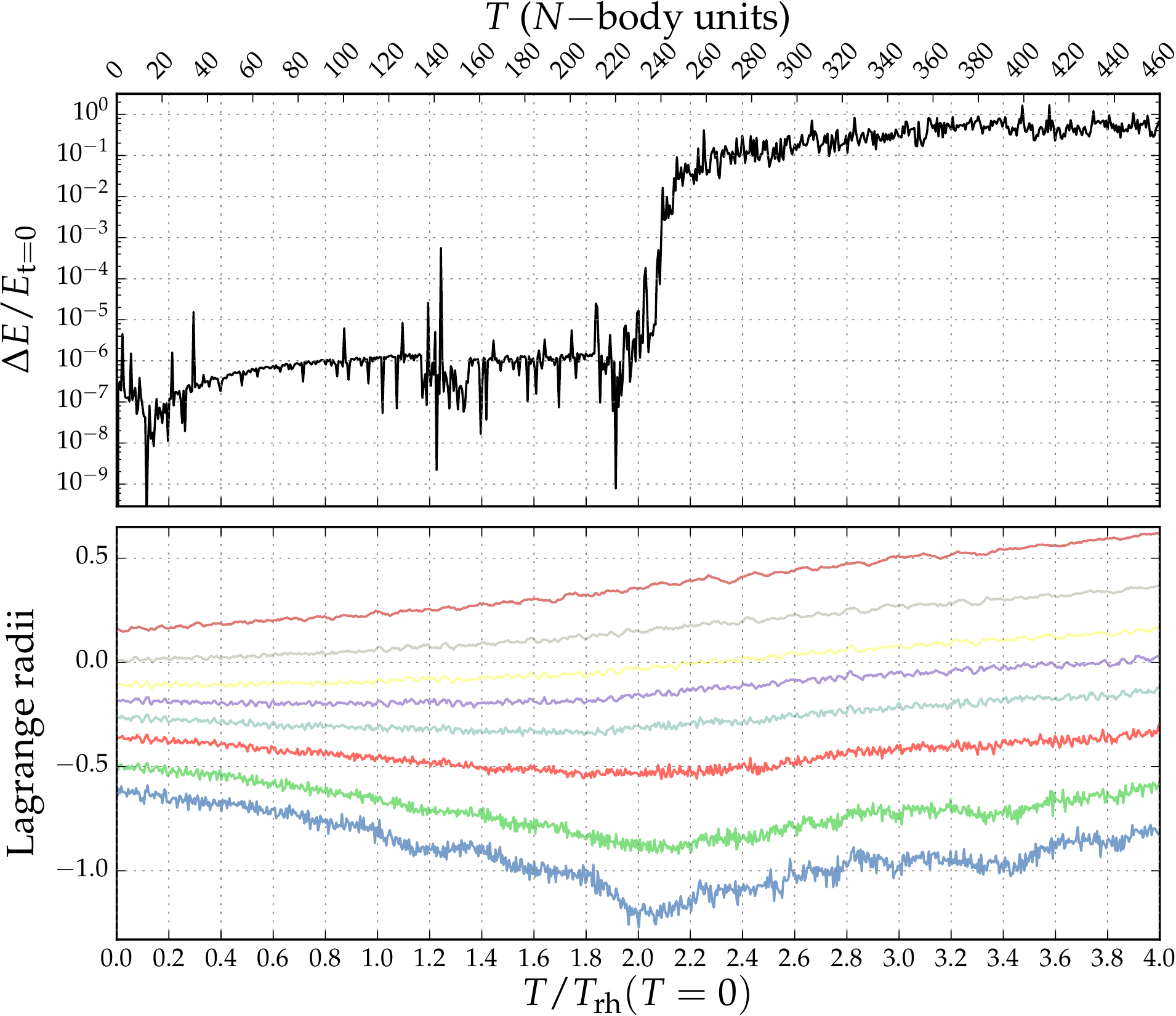}}
\caption
{
Plummer sphere using 8192 particles and following a Kroupa IMF.
{\it Top Panel:} Cumulative energy of the system.
{\it Bottom Panel:} Lagrange radii distribution for
    $0.5,\, 1,\, 2,\, 3,\, 4,\, 5,\, 6,\, 75$ and $90\%$ of the total mass.
}
\label{fig:kroupa_energy_radii}
\end{figure*}

\begin{figure*}
\resizebox{\hsize}{!}
          {\includegraphics[width=0.9\textwidth,clip]{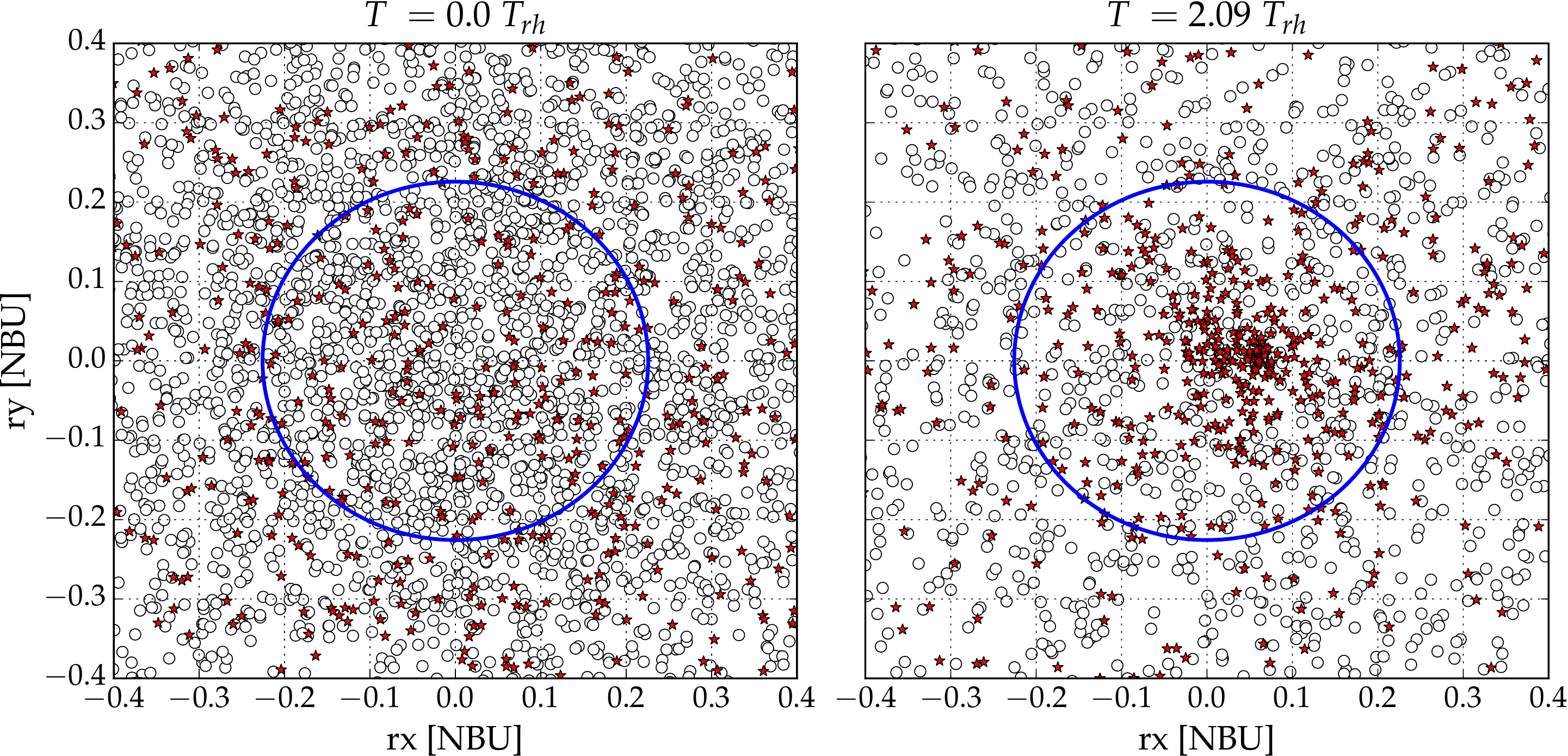}}
\caption
{
Inner section of a Plummer sphere with 8192 particles Plummer sphere following
a Kroupa IMF before (left) and after the core collapse (right) at $T=2.09\ T_{rh}$.
The blue circle depict the core radius at $T=0$.  The top $10\%$
of the heaviest particles in the system are marked as red stars, while all
other particles as empty circles.
}
\label{fig:kroupa_core_collapse}
\end{figure*}

Additionally to the single-mass calculations, we have also addressed multi-mass
systems. The fact of having an Initial Mass Function (IMF) accelerates the core
collapse of the system, as shown by many different authors
\citep{IW84,Spitzer87,KL97,KLG98}. In our calculations, we use a Plummer sphere with
a Kroupa IMF \citep{Kroupa01} and 8192 particles. In Figure~(\ref{fig:kroupa_energy_radii})
we present the evolution of the Lagrange radii and the energy conservation of the system.
We can see that the core collapse happens around $2\,T_{\rm rh}$, which is the point from
which the energy conservation becomes worse and worse, to achieve a value of about 6 orders
of magnitude worse than in phases before the collapse. Another way of depicting the collapse
is by identifying the heaviest $10\%$ of the stellar population and how it distributes in the
core radius as calculated at $T=0$. We can see this in Figure~(\ref{fig:kroupa_core_collapse}).

\begin{figure}
\resizebox{\hsize}{!}
          {\includegraphics[width=0.9\textwidth,clip]{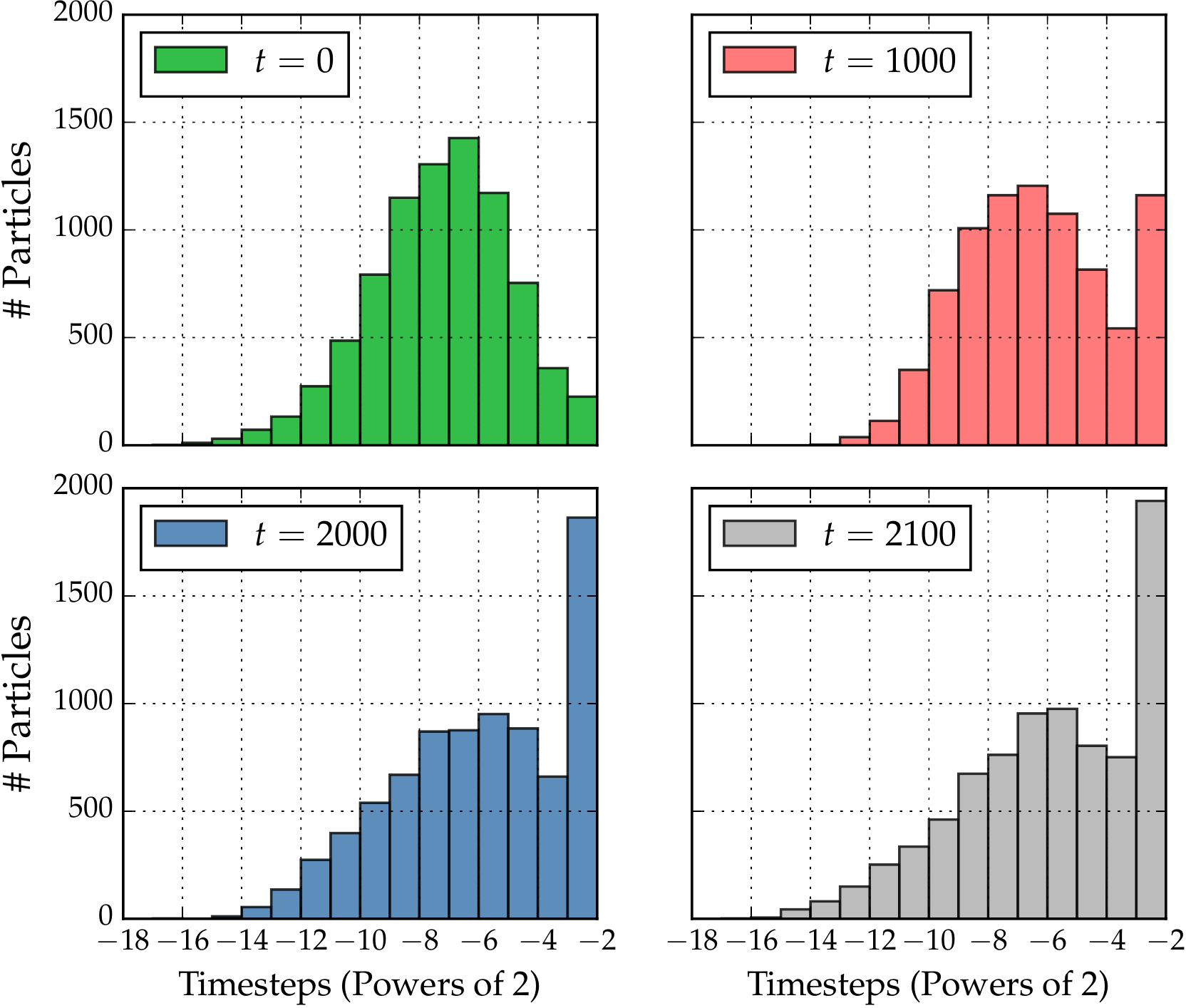}}
\caption
   {
Time step distribution of a Plummer sphere with $N=8192$ particles.
Four different times are shown,
(1) $t=0$ NBU, for an initial distribution (upper left panel)
(2) $t=1000$ NBU, a few half-mass relaxation times $\sim 9\, T_{\rm rh}$ (upper right panel),
(3) $t=2000$ NBU, a pre core-collapse stage with many particles leaving the core (lower left panel),
(4) $t=2100$ NBU, a post core-collapse stage with a few particles (mostly binaries)
reaching smaller time steps (lower right panel).
   }
\label{fig:dtdistribution}
\end{figure}

The equilibrium of the system can be evaluated by analysing the distribution of
the time steps.  As we have mentioned previously, in
Section~(\ref{sub:ImplScheme}), the initial distribution of time steps in the
system has a {log-normal} distribution, which in a balanced system must remain
similar, or close.
In Figure~(\ref{fig:dtdistribution}) we show the step
distribution after the core collapse for the single-mass system with $\epsilon
= 10^{-4}$

\section{Relativistic corrections}
\label{sec.PN}

{\GR} includes a treatment of relativistic orbits. This has been implemented
for the code to be able to study sources of gravitational waves. The approach
we have used is the post-Newtonian one, presented for the first time in an
$N-$body code in the work of \cite{KupiEtAl06} (and see also
\citealt{Amaro-SeoaneChen2016}) and later expanded to higher orders in
\cite{BremAmaro-SeoaneSpurzem2014}. The idea is to modify the accelerations in
the code to include relativistic corrections at 1PN, 2PN (periapsis shifts) and
2.5PN (energy loss in the form of gravitational wave emission). Contrary to the
scheme of \cite{KupiEtAl06}, which implements the modification in the
regularised binaries, in the case of {\GR}, the corrections are active for a
pair of two particles for which we set the softening to zero. The expressions
for the accelerations, as well as their time derivatives can be found in the
updated review of 2017 \cite{Amaro-SeoaneLRR2012}.

\begin{figure*}
\resizebox{\hsize}{!}
          {\includegraphics[scale=1,clip]{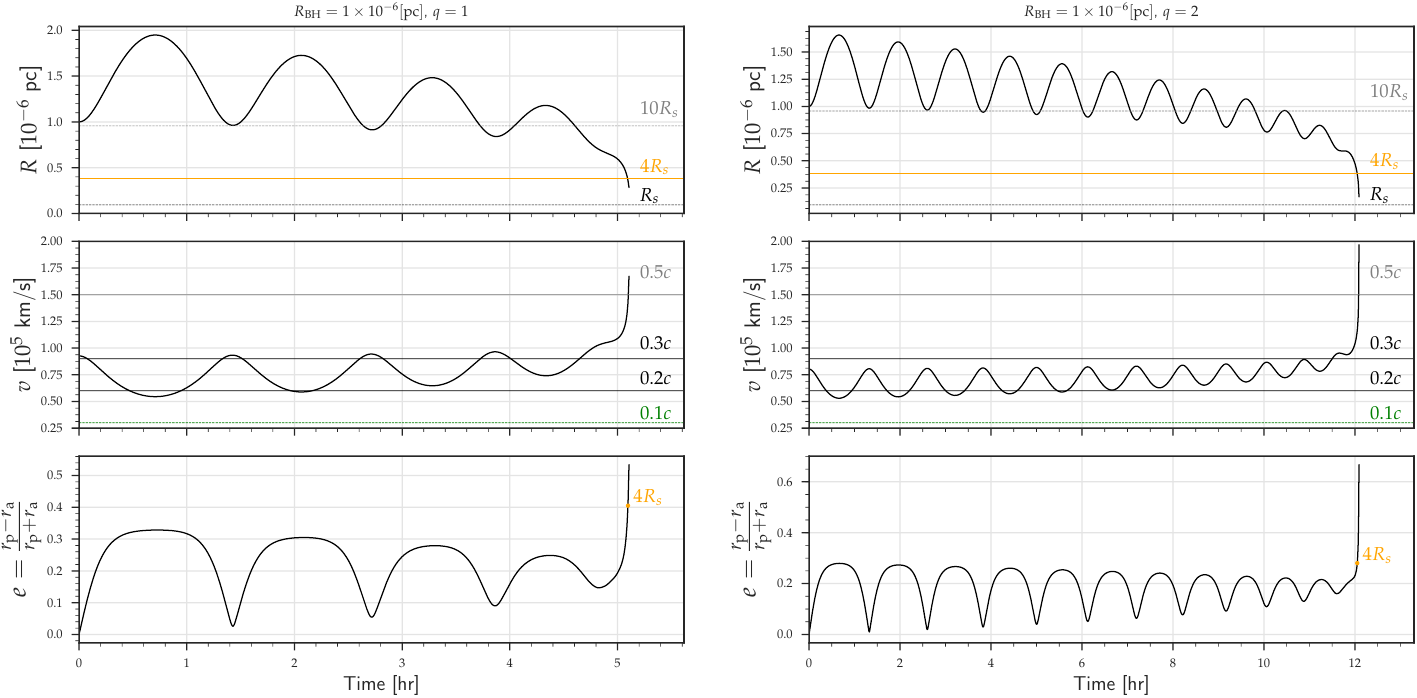}}
\caption
   {
\textit{Upper panel:} Evolution of the initial distance $R_{\rm BH}$
between the two super massive black holes for a mass ratio $q=1$ (and $q=2$ in the
set of three right panels). The
three horizontal lines correspond, from the top to the bottom, to
a distance of $10\,R_{\rm S}$, with $R_{\rm S}$ the Schwarzschild radius,
$4\,R_{\rm S}$, and $1\,R_{\rm S}$. \textit{Mid panel:} Evolution of the
relative velocity between the two super massive black holes. The four
horizontal lines correspond, from the top to the bottom, to a fraction of
the speed of light $c$ of $50\%$, $30\%$, $20\%$ and $10\%$. \textit{Bottom
panel:} Evolution of the eccentricity of the binary as a function of time in
hours. We mark the point in the evolution at which the separation is $4\,R_{\rm S}$
with an orange dot.
   }
\label{fig.PN1}
\end{figure*}

\begin{figure*}
\resizebox{\hsize}{!}
          {\includegraphics[scale=1,clip]{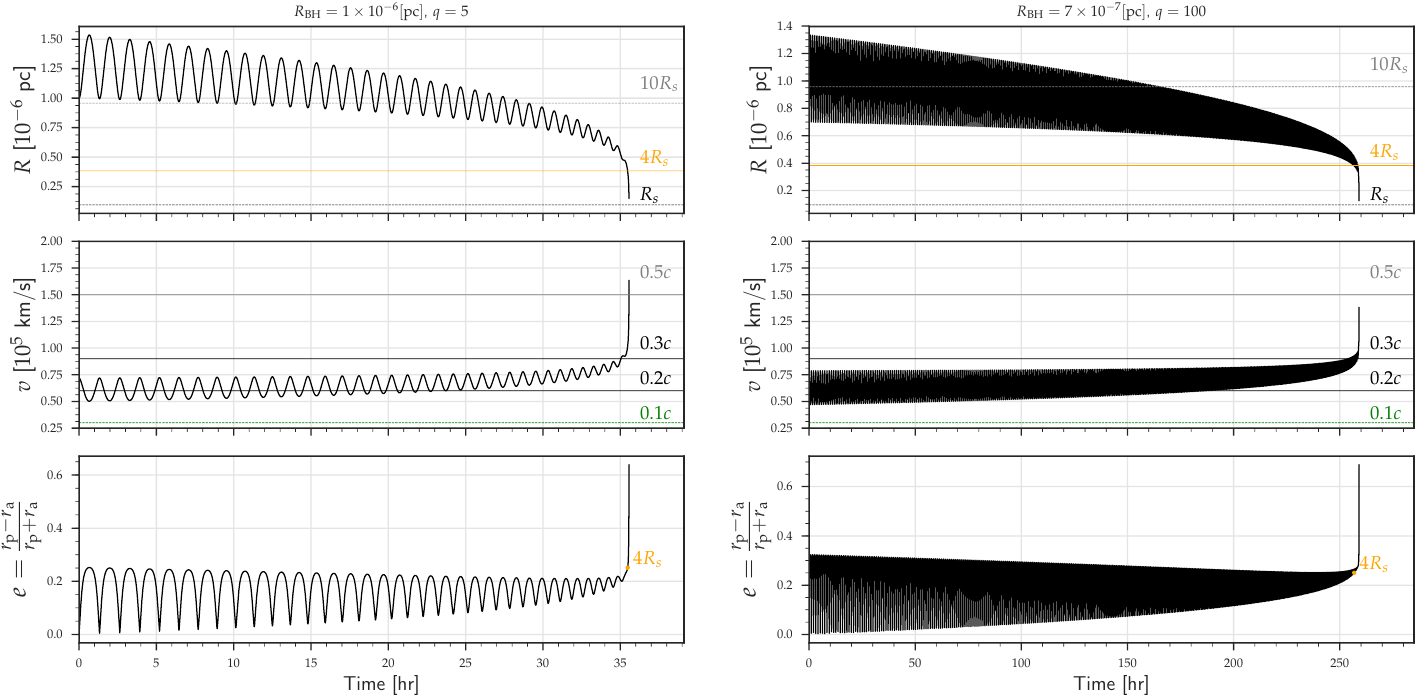}}
\caption
   {
Same as Fig.(\ref{fig.PN1}) but for mass ratios $q=5$ (left panels) and $q=100$ (right panels),
and a different initial separation, of $7\times 10^{-7}$ pc (also right panels).
   }
\label{fig.PN2}
\end{figure*}

We run a series of different tests for binaries with different mass ratios and
initial semi-major axis. In Fig.(\ref{fig.PN1}) we display the evolution of a
binary of two super massive black holes of total mass $1.33^6\,M_{\odot}$ and
mass ratios of 1 and 2.  In Fig.(\ref{fig.PN2}) we show mass ratios of 5 and 100,
and the latter starts with a smaller initial semi-major axis. For each of these
cases we plot the geometric distance, the relative velocity and the eccentricity. Higher
mass rations lead to a more complex structure in the evolution. We can see how the
relative velocity increases up to a significant fraction of the speed of light $c$
as the separation grows smaller. We however note that the post-Newtonian approach
should not be trusted for velocities larger than about $20\%\,c$.

\section{Conclusions and future work}
\label{sec.conclusions}

In this work we have presented the first version of our new {\nbody} code,
written purely in C/C++, using OpenMPI and CUDA, which we call {\GR}.  The
current version of our code provides an environment to evolve a
self-gravitating stellar system, and uses a H4 integration
scheme, using block time steps and softening, and features relativistic
corrections (periapsis shift and energy loss) for sources of gravitational
radiation. This first release of {\GR} has been mainly focused on users who can
have access to a machine hosting few GPUs, or usual parallel CPU systems.

We summarise here the main features of {\GR}:

\begin{enumerate}

    \item The code is written using an iterative and incremental development,
          which is methodology similar to the Assess, Parallelise, Optimise, Deploy
          (APOD) development cycle presented by NVIDIA.

    \item The code organisation is designed to be highly modular.
          Every critical process of the integrator is
          represented by a separate function or chain of functions.
          Our goal is to produce a code which can be read without
          difficulties, which makes easier future modifications or forks.

    \item Since maintainability is one of our main goals,
          the documentation is also a critical factor. We document
          every function in the inner procedure of the integrator.

    \item We use a H4 integrator scheme.

    \item The code uses block time steps to improve the performance of the
          integrator. We evolve particles in groups of block time steps, which
          allows for an update of several particles at the same time.

    \item We use GPU computing techniques, OpenMP and OpenMPI to parallelise
          the calculation of the gravitational interactions
          of our system after having localised the hot-spots of our algorithm.
          The main objective here was to be able to update a relatively small
          amount of particles which share a common time step in a given moment,
          a situation which is against the design of GPU cards, developed to
          reach a high parallelism.

\end{enumerate}

In this first release of {\GR} and first paper, we have presented a series of
classical tests of the code, as well as a study of the performance of its
different ``flavours'': the single CPU version, the MPI one and the
GPU version. We also address the role of the softening in the global evolution
of a system, as integrated with our code. As expected, the value of the
softening is crucial in determining the global dynamics, and should not be
taken lightly, in particular if one is interested in studying physical
phenomena for which relaxation is important, since using a softening translates
into {a maximum increase of the forces and the a smoothly declination
to zero}, which is approximate.
{To study a dynamical process, such as e.g. the collision of
two clusters, focusing on the short-term} (i.e. for times well below a
relaxation time) dynamical behaviour of the system, using a softening should be
fine, but the role of the parameter should be assessed carefully by exploring
different values.

The on-going development of {\GR} includes a close encounter solver, with a
time-symmetric integration scheme to treat binaries, such as the one presented
in the work of \cite{myriad}. Another immediate goal of the next releases, is
to include a central massive particle and the required corrections to the
gravitational forces so as to ensure a good conservation of the energy in the
system. This massive particle could be envisaged as a massive black hole in a
galactic centre or a star in a protoplanetary system. We also plan on bridging
the gap between spherical nucleus models that focus on collisional effects and
simulations of larger structure that are able to account for complex, more
realistic non-spherical geometry.
{Finally, a future goal is to include stellar evolution routines,
from which the modularity of our code will provide an easy scenario. One of the candidate modules for this
could be SEVN~\citep{SperaEtAl2015}.}

We will follow the APOD cycle presented in
this work, it is necessary to study new computational techniques, so as to
improve the performance of our code: from variable precision to new parallel
schemes to perform the force interaction calculation, using one or more GPU.

\appendix

\section{About the code}
\label{sec:compilation}

{\GR} is a C/C++ and CUDA application,
that uses the CUDA, OpenMPI and boost libraries.

As an overview, the compilation can be done with: {\tt make <flavour>},
for the {\tt cpu}, {\tt mpi} and {\tt gpu} versions.
A simple run of the code is displayed in the Listing~\ref{lst:code}.
\begin{tiny}
\begin{lstlisting}[float,
                   label=lst:code,
                   numbers=none,
                   language=make,
                   caption={Example run of the integrator. Columns, decimals, and information were modified
                   to fit the output on this document.}
                           ]
$ ./gravidy-gpu -i ../input/04-nbody-p1024_m1.in -p -t 1
[2017-01-28 01:60:56] [INFO] GPUs: 1
[2017-01-28 01:60:56] [INFO] Spl. 1024 particles in 1 GPUs
[2017-01-28 01:60:56] [INFO] GPU 0 particles: 1024
 Time  Iter  Nsteps     Energy      RelE      CumE     ETime
0.000     0       0  -2.56e-01  0.00e+00  0.00e+00  3.08e-02
0.125   698   30093  -2.56e-01  2.41e-07  2.41e-07  3.84e-01
0.250  1262   61319  -2.56e-01  1.10e-07  1.30e-07  6.60e-01
0.375  1897   91571  -2.56e-01  4.19e-08  8.84e-08  9.49e-01
0.500  2530  121963  -2.56e-01  8.51e-08  3.30e-09  1.23e+00
0.625  3132  150924  -2.56e-01  2.89e-08  3.23e-08  1.52e+00
0.750  3725  180446  -2.56e-01  1.39e-08  1.83e-08  1.76e+00
0.875  4354  212425  -2.56e-01  5.23e-07  5.41e-07  2.02e+00
1.000  5160  244165  -2.56e-01  2.32e-07  3.09e-07  2.32e+00
[2017-01-28 01:60:59] [SUCCESS] Finishing...
\end{lstlisting}
\end{tiny}

The URL hosting the project is \url{http://gravidy.xyz},
where you can find the prerequisites, how to get, compile and use the
code more detailed.
Additionally, documentation regarding the code, input and output files is included.

{Inside the repository, there is a {\tt scripts} directory with a set of classes
to be able to handle all the output files of the code.}

{The code was compiled using \texttt{gcc} (4.9.2), \texttt{openmpi}(1.6.5),
\texttt{CUDA}(6.0) and \texttt{boost} (1.55).}

{The following compilation FLAGs were used
\texttt{-O3 -Wall -fopenmp -pipe -fstack-protector -Wl,-z,relro -Wl,-z,now
-Wformat-security -Wpointer-arith -Wformat-nonliteral -Wl,-O1 -Wl,--discard-all
-Wl,--no-undefined -rdynamic}.}

\section{{\nbody} visualisation tool}
\label{sec:gravidyview}

\begin{figure}
\centering
\resizebox{0.9\hsize}{!}
          {\includegraphics[width=0.5\textwidth,clip]{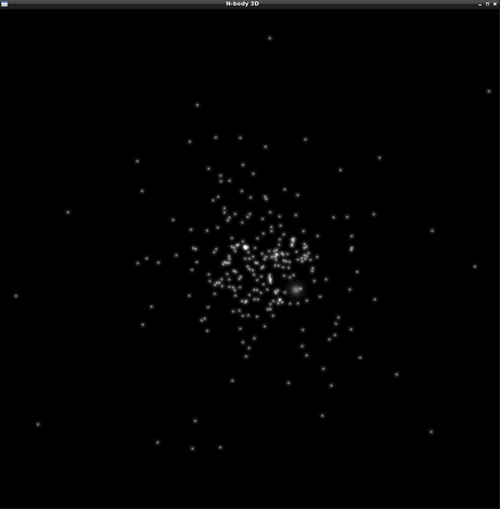}}
\caption
   {
   Snapshot pre-visualisation with {\tt GraviDyView} using a $N=1024$ system.
   }
\label{fig:gravidyview}
\end{figure}

A graphical representation of {\nbody} simulations is always an attractive
idea to display how the simulation was performed.
Due to this reason, we decided to write an small application to have a simple
3D visualisation of {\GR} snapshots, based in OpenGL.

{\sc GravidyView}
is a lightweight and simple OpenGL {\nbody} visualisation tool,
written in C/C++. It can be downloaded from:
\begin{itemize}
    \item \url{https://gitlab.com/cmaureir/gravidy-view}.
\end{itemize}

\section*{acknowledgements}
We are thankful for hints, help and discussion to Sverre Aarseth, Holger
Baumgardt, Peter Berczik, Douglas Heggie, Piet Hut, Simos Konstantinidis,
Patrick Brem, Keigo Nitadori, Ulf Löckmann and Rainer Spurzem.  In general,
we are indebted with the very friendly $N-$body community, for keeping their
codes publicly available to everybody.  PAS acknowledges support from the
Ram{\'o}n y Cajal Programme of the Ministry of Economy, Industry and
Competitiveness of Spain.  CMF thanks the support and guidance of Prof. Luis
Salinas during the development of his Master thesis, which was the motivation
of this project.  This work has been supported by the ``Dirección General de
Investigación y Postgrado'' (DGIP) by means of the ``Programa Incentivo a la
Iniciación Científica'' (PIIC) at the Universidad Técnica Federico Santa María,
the ``Comisión Nacional de Investigación
Científica y Tecnológica de Chile'' (CONICYT) trough its Master scholarships
program and the Transregio 7 ``Gravitational Wave Astronomy'' financed by the
Deutsche Forschungsgemeinschaft DFG (German Research Foundation). CMF
acknowledges support from the DFG Project  ``Supermassive black holes,
accretion discs, stellar dynamics and tidal disruptions'', awarded to PAS, and
the International Max-Planck Research School.


\begin{thebibliography}{}
\makeatletter
\relax
\def\mn@urlcharsother{\let\do\@makeother \do\$\do\&\do\#\do\^\do\_\do\%\do\~}
\def\mn@doi{\begingroup\mn@urlcharsother \@ifnextchar [ {\mn@doi@}
  {\mn@doi@[]}}
\def\mn@doi@[#1]#2{\def\@tempa{#1}\ifx\@tempa\@empty \href
  {http://dx.doi.org/#2} {doi:#2}\else \href {http://dx.doi.org/#2} {#1}\fi
  \endgroup}
\def\mn@eprint#1#2{\mn@eprint@#1:#2::\@nil}
\def\mn@eprint@arXiv#1{\href {http://arxiv.org/abs/#1} {{\tt arXiv:#1}}}
\def\mn@eprint@dblp#1{\href {http://dblp.uni-trier.de/rec/bibtex/#1.xml}
  {dblp:#1}}
\def\mn@eprint@#1:#2:#3:#4\@nil{\def\@tempa {#1}\def\@tempb {#2}\def\@tempc
  {#3}\ifx \@tempc \@empty \let \@tempc \@tempb \let \@tempb \@tempa \fi \ifx
  \@tempb \@empty \def\@tempb {arXiv}\fi \@ifundefined
  {mn@eprint@\@tempb}{\@tempb:\@tempc}{\expandafter \expandafter \csname
  mn@eprint@\@tempb\endcsname \expandafter{\@tempc}}}

\bibitem[\protect\citeauthoryear{{Aarseth}}{{Aarseth}}{1985}]{Aarseth85}
{Aarseth} S.~J.,  1985, in {Brackbill} J.~U.,  {Cohen} B.~I.,  eds, Multiple
  time scales, p. 377 - 418. pp 377--418

\bibitem[\protect\citeauthoryear{{Aarseth}}{{Aarseth}}{1999}]{Aarseth99}
{Aarseth} S.~J.,  1999, The Publications of the Astronomical Society of the
  Pacific, \href
  {http://adsabs.harvard.edu/cgi-bin/nph-bib_query?bibcode=1999PASP..111.1333A&db_key=AST}
  {111, 1333}

\bibitem[\protect\citeauthoryear{{Aarseth}}{{Aarseth}}{2003}]{Aarseth03}
{Aarseth} S.~J.,  2003, {Gravitational N-Body Simulations}.
ISBN 0521432723.~Cambridge, UK: Cambridge University Press, November 2003.

\bibitem[\protect\citeauthoryear{{Aarseth} \& {Zare}}{{Aarseth} \&
  {Zare}}{1974}]{AarsethZare74}
{Aarseth} S.~J.,  {Zare} K.,  1974, \mn@doi [Celestial Mechanics]
  {10.1007/BF01227619}, \href
  {http://adsabs.harvard.edu/abs/1974CeMec..10..185A} {10, 185}

\bibitem[\protect\citeauthoryear{{Aarseth}, {Henon}  \& {Wielen}}{{Aarseth}
  et~al.}{1974}]{Aarseth74}
{Aarseth} S.~J.,  {Henon} M.,   {Wielen} R.,  1974, A\&A, \href
  {http://adsabs.harvard.edu/abs/1974A%26A....37..183A} {37, 183}

\bibitem[\protect\citeauthoryear{{Amaro-Seoane}}{{Amaro-Seoane}}{2012}]{Amaro-SeoaneLRR2012}
{Amaro-Seoane} P.,  2012, preprint, \href
  {http://adsabs.harvard.edu/abs/2012arXiv1205.5240A} {} (\mn@eprint {arXiv}
  {1205.5240})

\bibitem[\protect\citeauthoryear{{Amaro-Seoane} \& {Chen}}{{Amaro-Seoane} \&
  {Chen}}{2016}]{Amaro-SeoaneChen2016}
{Amaro-Seoane} P.,  {Chen} X.,  2016, \mn@doi [MNRAS] {10.1093/mnras/stw503},
  \href {http://adsabs.harvard.edu/abs/2016MNRAS.458.3075A} {458, 3075}

\bibitem[\protect\citeauthoryear{{Amaro-Seoane}, {Freitag}  \&
  {Spurzem}}{{Amaro-Seoane} et~al.}{2004}]{ASEtAl04}
{Amaro-Seoane} P.,  {Freitag} M.,   {Spurzem} R.,  2004, MNRAS, \href
  {http://adsabs.harvard.edu/cgi-bin/nph-bib_query?bibcode=2004astro.ph..1163A&amp;db_key=PRE}
  {}

\bibitem[\protect\citeauthoryear{{Barnes} \& {Hut}}{{Barnes} \&
  {Hut}}{1986}]{BarnesHut86}
{Barnes} J.,  {Hut} P.,  1986, \mn@doi [Nat] {10.1038/324446a0}, \href
  {http://adsabs.harvard.edu/abs/1986Natur.324..446B} {324, 446}

\bibitem[\protect\citeauthoryear{{Belleman}, {B{\'e}dorf}  \& {Portegies
  Zwart}}{{Belleman} et~al.}{2008}]{Belleman2008}
{Belleman} R.~G.,  {B{\'e}dorf} J.,   {Portegies Zwart} S.~F.,  2008, \mn@doi
  [New Astronomy] {10.1016/j.newast.2007.07.004}, \href
  {http://adsabs.harvard.edu/abs/2008NewA...13..103B} {13, 103}

\bibitem[\protect\citeauthoryear{{Berczik} et~al.,}{{Berczik}
  et~al.}{2011}]{berczik2011high}
{Berczik} P.,  et~al., 2011, \href
  {http://adsabs.harvard.edu/abs/2011hpc..conf....8B} {pp 8--18}

\bibitem[\protect\citeauthoryear{{Berczik}, {Spurzem}, {Wang}, {Zhong}  \&
  {Huang}}{{Berczik} et~al.}{2013}]{bercziketal2013}
{Berczik} P.,  {Spurzem} R.,  {Wang} L.,  {Zhong} S.,   {Huang} S.,  2013, in
  Third International Conference ''High Performance Computing'', HPC-UA 2013,
  p. 52-59. pp 52--59 (\mn@eprint {arXiv} {1312.1789})

\bibitem[\protect\citeauthoryear{{Brem}, {Amaro-Seoane}  \& {Spurzem}}{{Brem}
  et~al.}{2013}]{BremAmaro-SeoaneSpurzem2014}
{Brem} P.,  {Amaro-Seoane} P.,   {Spurzem} R.,  2013, \mn@doi [MNRAS]
  {10.1093/mnras/stt1220}, \href
  {http://adsabs.harvard.edu/abs/2013MNRAS.434.2999B} {434, 2999}

\bibitem[\protect\citeauthoryear{{Capuzzo-Dolcetta} \&
  {Spera}}{{Capuzzo-Dolcetta} \& {Spera}}{2013}]{CapuzzoDolcettaSpera2013}
{Capuzzo-Dolcetta} R.,  {Spera} M.,  2013, \mn@doi [Computer Physics
  Communications] {10.1016/j.cpc.2013.07.005}, \href
  {http://adsabs.harvard.edu/abs/2013CoPhC.184.2528C} {184, 2528}

\bibitem[\protect\citeauthoryear{{Capuzzo-Dolcetta}, {Spera}  \&
  {Punzo}}{{Capuzzo-Dolcetta} et~al.}{2013}]{Capuzzo-DolcettaEtAl2013}
{Capuzzo-Dolcetta} R.,  {Spera} M.,   {Punzo} D.,  2013, \mn@doi [Journal of
  Computational Physics] {10.1016/j.jcp.2012.11.013}, \href
  {http://adsabs.harvard.edu/abs/2013JCoPh.236..580C} {236, 580}

\bibitem[\protect\citeauthoryear{{Fukushige}, {Makino}  \& {Kawai}}{{Fukushige}
  et~al.}{2005}]{GRAPE6A}
{Fukushige} T.,  {Makino} J.,   {Kawai} A.,  2005, PASJ, \href
  {http://adsabs.harvard.edu/cgi-bin/nph-bib_query?bibcode=2005PASJ...57.1009F&db_key=AST}
  {57, 1009}

\bibitem[\protect\citeauthoryear{{Gaburov}, {Harfst}  \& {Zwart}}{{Gaburov}
  et~al.}{2009}]{Gaburov2009}
{Gaburov} E.,  {Harfst} S.,   {Zwart} S.~P.,  2009, \mn@doi [New Astronomy]
  {10.1016/j.newast.2009.03.002}, \href
  {http://adsabs.harvard.edu/abs/2009NewA...14..630G} {14, 630}

\bibitem[\protect\citeauthoryear{{Giersz} \& {Spurzem}}{{Giersz} \&
  {Spurzem}}{1994}]{Giersz94}
{Giersz} M.,  {Spurzem} R.,  1994, MNRAS, \href
  {http://adsabs.harvard.edu/abs/1994MNRAS.269..241G} {269, 241}

\bibitem[\protect\citeauthoryear{Greendard}{Greendard}{1987}]{GreendardThesis}
Greendard L.,  1987, PhD thesis, Yale University, New Haven, CT

\bibitem[\protect\citeauthoryear{{Hamada} \& {Iitaka}}{{Hamada} \&
  {Iitaka}}{2007}]{Hamada2007}
{Hamada} T.,  {Iitaka} T.,  2007, New Astronomy, \href
  {http://adsabs.harvard.edu/abs/2007astro.ph..3100H} {}

\bibitem[\protect\citeauthoryear{{Harfst}, {Gualandris}, {Merritt}  \&
  {Mikkola}}{{Harfst} et~al.}{2008}]{harfst2008}
{Harfst} S.,  {Gualandris} A.,  {Merritt} D.,   {Mikkola} S.,  2008, \mn@doi
  [MNRAS] {10.1111/j.1365-2966.2008.13557.x}, \href
  {http://adsabs.harvard.edu/abs/2008MNRAS.389....2H} {389, 2}

\bibitem[\protect\citeauthoryear{{Heggie} \& {Hut}}{{Heggie} \&
  {Hut}}{2003}]{HeggieHut03}
{Heggie} D.,  {Hut} P.,  2003, {The Gravitational Million-Body Problem: A
  Multidisciplinary Approach to Star Cluster Dynamics, by Douglas Heggie and
  Piet Hut.~ Cambridge University Press, 2003, 372 pp.}

\bibitem[\protect\citeauthoryear{{Heggie} \& {Mathieu}}{{Heggie} \&
  {Mathieu}}{1986}]{Heggie1986}
{Heggie} D.~C.,  {Mathieu} R.~D.,  1986, in {Hut} P.,  {McMillan} S.~L.~W.,
  eds,  Lecture Notes in Physics, Berlin Springer Verlag Vol. 267, The Use of
  Supercomputers in Stellar Dynamics. p.~233, \mn@doi{10.1007/BFb0116419}

\bibitem[\protect\citeauthoryear{{H{\'e}non}}{{H{\'e}non}}{1971}]{Henon}
{H{\'e}non} M.~H.,  1971, \mn@doi [A\&AS] {10.1007/BF00649201}, \href
  {http://adsabs.harvard.edu/abs/1971Ap%26SS..14..151H} {14, 151}

\bibitem[\protect\citeauthoryear{{Holmberg}}{{Holmberg}}{1941}]{Holmberg1941}
{Holmberg} E.,  1941, \mn@doi [ApJ] {10.1086/144344}, \href
  {http://adsabs.harvard.edu/abs/1941ApJ....94..385H} {94, 385}

\bibitem[\protect\citeauthoryear{{Hut}}{{Hut}}{2003}]{hut2003}
{Hut} P.,  2003, in {Makino} J.,  {Hut} P.,  eds,  IAU Symposium Vol. 208,
  Astrophysical Supercomputing using Particle Simulations. p.~331 (\mn@eprint
  {} {arXiv:astro-ph/0204431})

\bibitem[\protect\citeauthoryear{{Inagaki} \& {Wiyanto}}{{Inagaki} \&
  {Wiyanto}}{1984}]{IW84}
{Inagaki} S.,  {Wiyanto} P.,  1984, PASJ, 36, 391

\bibitem[\protect\citeauthoryear{{Kim} \& {Lee}}{{Kim} \& {Lee}}{1997}]{KL97}
{Kim} S.~S.,  {Lee} H.~M.,  1997, Journal of Korean Astronomical Society, 30,
  115

\bibitem[\protect\citeauthoryear{{Kim}, {Lee}  \& {Goodman}}{{Kim}
  et~al.}{1998}]{KLG98}
{Kim} S.~S.,  {Lee} H.~M.,   {Goodman} J.,  1998, ApJ, \href
  {http://adsabs.harvard.edu/cgi-bin/nph-bib_query?bibcode=1998ApJ...495..786K&amp;db_key=AST}
  {495, 786}

\bibitem[\protect\citeauthoryear{{Konstantinidis} \&
  {Kokkotas}}{{Konstantinidis} \& {Kokkotas}}{2010}]{myriad}
{Konstantinidis} S.,  {Kokkotas} K.~D.,  2010, \mn@doi [A\&A]
  {10.1051/0004-6361/200913890}, \href
  {http://adsabs.harvard.edu/abs/2010A%26A...522A..70K} {522, A70}

\bibitem[\protect\citeauthoryear{{Kroupa}}{{Kroupa}}{2001}]{Kroupa01}
{Kroupa} P.,  2001, MNRAS, \href
  {http://adsabs.harvard.edu/cgi-bin/nph-bib_query?bibcode=2001MNRAS.322..231K&amp;db_key=AST}
  {322, 231}

\bibitem[\protect\citeauthoryear{{Kupi}, {Amaro-Seoane}  \& {Spurzem}}{{Kupi}
  et~al.}{2006}]{KupiEtAl06}
{Kupi} G.,  {Amaro-Seoane} P.,   {Spurzem} R.,  2006, \mn@doi [MNRAS]
  {10.1111/j.1745-3933.2006.00205.x}, \href
  {http://adsabs.harvard.edu/cgi-bin/nph-bib_query?bibcode=2006MNRAS.tmpL..77K&db_key=AST}
  {pp~L77+}

\bibitem[\protect\citeauthoryear{{K{\"u}pper}, {Maschberger}, {Kroupa}  \&
  {Baumgardt}}{{K{\"u}pper} et~al.}{2011}]{mcluster}
{K{\"u}pper} A.~H.~W.,  {Maschberger} T.,  {Kroupa} P.,   {Baumgardt} H.,
  2011, \mn@doi [MNRAS] {10.1111/j.1365-2966.2011.19412.x}, \href
  {http://adsabs.harvard.edu/abs/2011MNRAS.417.2300K} {417, 2300}

\bibitem[\protect\citeauthoryear{{Kustaanheimo} \& {Stiefel}}{{Kustaanheimo} \&
  {Stiefel}}{1965}]{KS65}
{Kustaanheimo} P.~E.,  {Stiefel} E.~L.,  1965, J. Reine Angew. Math., 218, 204

\bibitem[\protect\citeauthoryear{{Makino}}{{Makino}}{1991}]{Makino91}
{Makino} J.,  1991, \mn@doi [ApJ] {10.1086/169751}, \href
  {http://adsabs.harvard.edu/abs/1991ApJ...369..200M} {369, 200}

\bibitem[\protect\citeauthoryear{{Makino}}{{Makino}}{1998}]{Makino98}
{Makino} J.,  1998, Highlights in Astronomy, 11, 597

\bibitem[\protect\citeauthoryear{{Makino} \& {Aarseth}}{{Makino} \&
  {Aarseth}}{1992}]{ma92}
{Makino} J.,  {Aarseth} S.~J.,  1992, PASJ, \href
  {http://adsabs.harvard.edu/abs/1992PASJ...44..141M} {44, 141}

\bibitem[\protect\citeauthoryear{{Makino} \& {Taiji}}{{Makino} \&
  {Taiji}}{1998}]{MT98}
{Makino} J.,  {Taiji} M.,  1998, {Scientific simulations with special-purpose
  computers : The GRAPE systems}.
Scientific simulations with special-purpose computers : The GRAPE systems /by
  Junichiro Makino \& Makoto Taiji.~Chichester ; Toronto : John Wiley \& Sons,
  c1998.

\bibitem[\protect\citeauthoryear{Nguyen}{Nguyen}{2007}]{gpuGems3}
Nguyen H.,  2007, Gpu gems 3, first edn.
Addison-Wesley Professional

\bibitem[\protect\citeauthoryear{Nitadori}{Nitadori}{2009}]{keigo}
Nitadori K.,  2009, PhD thesis, University of Tokyo

\bibitem[\protect\citeauthoryear{{Nitadori} \& {Aarseth}}{{Nitadori} \&
  {Aarseth}}{2012}]{NitadoriAarseth2012}
{Nitadori} K.,  {Aarseth} S.~J.,  2012, \mn@doi [MNRAS]
  {10.1111/j.1365-2966.2012.21227.x}, \href
  {http://adsabs.harvard.edu/abs/2012MNRAS.424..545N} {424, 545}

\bibitem[\protect\citeauthoryear{{Nitadori} \& {Makino}}{{Nitadori} \&
  {Makino}}{2008}]{Nitadori6th8th}
{Nitadori} K.,  {Makino} J.,  2008, \mn@doi [na]
  {10.1016/j.newast.2008.01.010}, \href
  {http://adsabs.harvard.edu/abs/2008NewA...13..498N} {13, 498}

\bibitem[\protect\citeauthoryear{{Plummer}}{{Plummer}}{1911}]{plummer1911}
{Plummer} H.~C.,  1911, MNRAS, \href
  {http://adsabs.harvard.edu/abs/1911MNRAS..71..460P} {71, 460}

\bibitem[\protect\citeauthoryear{{Portegies Zwart}, {McMillan}, {Hut}  \&
  {Makino}}{{Portegies Zwart} et~al.}{2001a}]{PortegiesZwartEtAl01}
{Portegies Zwart} S.~F.,  {McMillan} S.~L.~W.,  {Hut} P.,   {Makino} J.,
  2001a, MNRAS, \href {http://adsabs.harvard.edu/abs/2001MNRAS.321..199P} {321,
  199}

\bibitem[\protect\citeauthoryear{{Portegies Zwart}, {McMillan}, {Hut}  \&
  {Makino}}{{Portegies Zwart} et~al.}{2001b}]{portegies2001}
{Portegies Zwart} S.~F.,  {McMillan} S.~L.~W.,  {Hut} P.,   {Makino} J.,
  2001b, MNRAS, \href {http://adsabs.harvard.edu/abs/2001MNRAS.321..199P} {321,
  199}

\bibitem[\protect\citeauthoryear{{Portegies Zwart}, {Belleman}  \&
  {Geldof}}{{Portegies Zwart} et~al.}{2007}]{Portegies2007a}
{Portegies Zwart} S.~F.,  {Belleman} R.~G.,   {Geldof} P.~M.,  2007, \mn@doi
  [New Astronomy] {10.1016/j.newast.2007.05.004}, \href
  {http://adsabs.harvard.edu/abs/2007NewA...12..641P} {12, 641}

\bibitem[\protect\citeauthoryear{{Press}}{{Press}}{1986}]{Press86}
{Press} W.~H.,  1986, in {Hut} P.,  {McMillan} S. L.~W.,  eds, The Use of
  Supercomputers in Stellar Dynamics. Springer-Verlag, p.~184

\bibitem[\protect\citeauthoryear{{Schneider}, {Amaro-Seoane}  \&
  {Spurzem}}{{Schneider} et~al.}{2011}]{SchneiderEtAl11}
{Schneider} J.,  {Amaro-Seoane} P.,   {Spurzem} R.,  2011, \mn@doi [MNRAS]
  {10.1111/j.1365-2966.2010.17454.x}, \href
  {http://adsabs.harvard.edu/abs/2011MNRAS.410..432S} {410, 432}

\bibitem[\protect\citeauthoryear{{Spera}, {Mapelli}  \& {Bressan}}{{Spera}
  et~al.}{2015}]{SperaEtAl2015}
{Spera} M.,  {Mapelli} M.,   {Bressan} A.,  2015, \mn@doi [MNRAS]
  {10.1093/mnras/stv1161}, \href
  {http://adsabs.harvard.edu/abs/2015MNRAS.451.4086S} {451, 4086}

\bibitem[\protect\citeauthoryear{{Spitzer}}{{Spitzer}}{1987}]{Spitzer87}
{Spitzer} L.,  1987, {Dynamical evolution of globular clusters}.
Princeton, NJ, Princeton University Press, 1987, 191 p.

\bibitem[\protect\citeauthoryear{{Spitzer} \& {Hart}}{{Spitzer} \&
  {Hart}}{1971}]{SH71b}
{Spitzer} L.~J.,  {Hart} M.~H.,  1971, ApJ, 166, 483

\bibitem[\protect\citeauthoryear{{Spurzem}}{{Spurzem}}{1999}]{Spurzem1999}
{Spurzem} R.,  1999, Journal of Computational and Applied Mathematics, \href
  {http://adsabs.harvard.edu/abs/1999JCoAM.109..407S} {109, 407}

\bibitem[\protect\citeauthoryear{{Taiji}, {Makino}, {Fukushige}, {Ebisuzaki}
  \& {Sugimoto}}{{Taiji} et~al.}{1996}]{TMFES96}
{Taiji} M.,  {Makino} J.,  {Fukushige} T.,  {Ebisuzaki} T.,   {Sugimoto} D.,
  1996, in {Hut} P.,  {Makino} J.,  eds, IAU Symp. 174: Dynamical Evolution of
  Star Clusters: Confrontation of Theory and Observations. p.~141

\bibitem[\protect\citeauthoryear{{Wang}, {Spurzem}, {Aarseth}, {Nitadori},
  {Berczik}, {Kouwenhoven}  \& {Naab}}{{Wang} et~al.}{2015}]{WangEtAl2015}
{Wang} L.,  {Spurzem} R.,  {Aarseth} S.,  {Nitadori} K.,  {Berczik} P.,
  {Kouwenhoven} M.~B.~N.,   {Naab} T.,  2015, \mn@doi [mn]
  {10.1093/mnras/stv817}, \href
  {http://adsabs.harvard.edu/abs/2015MNRAS.450.4070W} {450, 4070}

\bibitem[\protect\citeauthoryear{{Wang} et~al.,}{{Wang}
  et~al.}{2016}]{WangEtAl2016}
{Wang} L.,  et~al., 2016, \mn@doi [mn] {10.1093/mnras/stw274}, \href
  {http://adsabs.harvard.edu/abs/2016MNRAS.458.1450W} {458, 1450}

\bibitem[\protect\citeauthoryear{{von Hoerner}}{{von
  Hoerner}}{1960}]{vonHoerner1960}
{von Hoerner} S.,  1960, Z. Astrophys., \href
  {http://adsabs.harvard.edu/abs/1960ZA.....50..184V} {50, 184}

\bibitem[\protect\citeauthoryear{{von Hoerner}}{{von
  Hoerner}}{1963}]{vonHoerner1963}
{von Hoerner} S.,  1963, Z. Astrophys., \href
  {http://adsabs.harvard.edu/abs/1963ZA.....57...47V} {57, 47}

\makeatother
\end{thebibliography}
\end{document}